\documentclass[twocolumn,twocolappendix]{aastex701}
\usepackage{color}
\usepackage[titletoc]{appendix}
\usepackage{amsmath}
\usepackage{amssymb}
\usepackage{mathtools}
\usepackage{upgreek}
\usepackage{float}
\usepackage{comment}
\usepackage{enumitem}
\usepackage{natbib}
\usepackage{graphicx}
\usepackage{bm}
\usepackage{totcount}
\usepackage{multirow}
\usepackage{pifont}
\usepackage{afterpage}

\usepackage{xcolor}

\newtotcounter{citnum} 
\def\oldbibitem{} \let\oldbibitem=\bibitem
\def\bibitem{\stepcounter{citnum}\oldbibitem}

\shortauthors{Sethi et. al.}
\shorttitle{Occurrence Rates of Close-in Sub-Neptunes}
\begin{document} 
\title{Coupled Orbital and Interior Evolution of Sub-Neptunes}
\author[0000-0002-6576-3346]{Ritika Sethi}
\affiliation{Department of Physics, Massachusetts Institute of Technology, Cambridge, MA 02139, USA}
\affiliation{MIT Kavli Institute for Astrophysics and Space Research, Massachusetts Institute of Technology, Cambridge, MA 02139, USA}
\email{rsethi@mit.edu}

\author[0000-0003-3130-2282]{Sarah Millholland}
\affiliation{Department of Physics, Massachusetts Institute of Technology, Cambridge, MA 02139, USA}
\affiliation{MIT Kavli Institute for Astrophysics and Space Research, Massachusetts Institute of Technology, Cambridge, MA 02139, USA}
\email{sarah.millholland@mit.edu}

\author[0000-0003-4992-8427]{Tim Hallatt}
\affiliation{Department of Physics, Massachusetts Institute of Technology, Cambridge, MA 02139, USA}
\affiliation{MIT Kavli Institute for Astrophysics and Space Research, Massachusetts Institute of Technology, Cambridge, MA 02139, USA}
\email{email}

\begin{abstract}
Recent observations have yielded the first measurements of young exoplanet demographics. Close-in sub-Neptune occurrence rates appear to rise from young ($10-100$ Myr) to intermediate (100 Myr $-$ 1 Gyr) ages and then decline sharply in the old ($\gtrsim 1$ Gyr) Kepler field population. In this paper, we test whether these observations can be explained by the effects of planetary cooling, atmospheric mass loss, and tidal orbital migration, which we model through a fully coupled evolution framework. 
The orbital evolution is assumed to operate exclusively through high-eccentricity migration, in effort to estimate the maximum possible contribution from this migration channel. In reality, only a subset of planetary systems are expected to undergo HEM.
We find that high-eccentricity migration rapidly populates the close-in sub-Neptune bin, producing a sharp rise in occurrence within the first $\sim 15$ Myr. After this early phase, the occurrence evolves only weakly. 
While the modeled 
young to intermediate-age evolution is broadly consistent with the observed trend within the uncertainty limits, it does not support the idea of a sustained young-to-intermediate rise driven by tidal migration. It also fails to reproduce the sharp decline in occurrence seen from intermediate to old ages, motivating additional physics or formation channels to be included in future models.

\end{abstract}

\section{Introduction} \label{sec: Introduction} 
Close-in small planets with sizes between Earth and Neptune are the most common planets revealed by \textit{Kepler}, K2, and TESS \citep[e.g.,][]{2012ApJS..201...15H,2013ApJ...766...81F,2013ApJ...770...69P}. This population is separated into super-Earths and sub-Neptunes by the observed radius valley, a scarcity of planets near $\sim 1.5$--$2~R_\oplus$ \citep{2013ApJ...775..105O,2017ApJ...847...29O,2017AJ....154..109F,2018MNRAS.479.4786V}. The valley is commonly thought to reflect a transition between planets that retain substantial primordial H/He envelopes and appear as sub-Neptunes and smaller planets with little or no H/He envelope, which may either be stripped cores \citep{2017ApJ...847...29O,2018MNRAS.476..759G,2019MNRAS.487...24G} or planets that formed with only thin atmospheres \citep{2015A&A...578A..36O, 2017MNRAS.470.1750I, 2021ApJ...908...32L}. 

Although close-in planet demographic features are well established for the mature \textit{Kepler} field population, tracing how they emerge with age remains observationally challenging, especially for small planets around young stars exhibiting high spot coverage and frequent flares \citep{2008ApJ...687.1264M, 2020MNRAS.496.1197B, 2020AJ....160..219F}. Recent observational progress, however, is making it possible to study the age evolution of close-in sub-Neptunes at a population level. Such measurements provide a way to directly observe the physical processes responsible for sculpting the mature population. Occurrence rates of close-in sub-Neptunes have been measured for young and intermediate age stars using TESS and K2 \citep{2023AJ....166..175F, 2023AJ....166..248C,2024AJ....167..210V,2025AJ....169..208F}. 

In particular, \citet{2025AJ....169..208F} studied a sample of FGK stars divided into young ($10$--$100$~Myr), intermediate ($100$~Myr--$1$~Gyr), and old ($\gtrsim~1$~Gyr) ages, and measured the occurrence rates of close-in sub-Neptunes across these age ranges. They defined close-in sub-Neptunes as planets with orbital periods $P_{\rm orb} < 12$ days and radii $1.8~R_\oplus<R_p<10~R_\oplus$. For consistency with their work, we adopt the same definition throughout this paper. We also define close-in super-Earths as planets with $P_{\rm orb}<12$ days and $1~R_\oplus<R_p<1.8~R_\oplus$. Hereafter, we refer to these close-in planets simply as sub-Neptunes and super-Earths unless stated otherwise.

\citet{2025AJ....169..208F} found that the sub-Neptune occurrence rate increases from young to intermediate ages (but is flat within $2\sigma$ uncertainties) and then declines significantly toward the old ages represented by the \textit{Kepler} field population. They hypothesized that the early rise, if present, could be driven by tidal migration, while the later decline could reflect atmospheric mass-loss removing planets from the sub-Neptune radius range. 

In this paper, we aim to test the hypotheses of \citet{2025AJ....169..208F} by modeling three key physical processes that could drive the age evolution of small planet occurrence rates. (1) Thermal cooling causes young planets that form with high internal entropy and inflated radii to contract as they age \citep{2012ApJ...761...59L,2014ApJ...792....1L,2026ApJ...997..138H}. This radius evolution can move planets into or out of the sub-Neptune bin even at fixed mass. (2) Atmospheric mass loss erodes irradiated planets, reducing their radii and in some cases transforming sub-Neptunes into super-Earths \citep[e.g.,][]{2013ApJ...775..105O, 2013ApJ...776....2L, 2017ApJ...847...29O}, thereby decreasing the occurrence rate within the sub-Neptune bin. (3) Tidal migration can deliver planets into the close-in bin. In particular, planets excited to high eccentricities through dynamical processes can undergo tidal dissipation and circularize at short orbital periods in a process known as high-eccentricity migration (HEM) \citep[][]{1980A&A....92..167H, 2026A&A...709L..17C}. 

Although these processes have often been studied separately, their combined effect on population-level occurrence rates remains less well explored. Moreover, they are intrinsically coupled. The planetary radius, set by thermal cooling and atmospheric mass loss, affects tidal dissipation rates. Tidal migration changes the orbital separation and incident stellar flux, which regulate mass loss and thermal evolution. Atmospheric loss changes the envelope mass fraction and radius, feeding back onto both cooling and tidal evolution. A coupled treatment is therefore needed to accurately predict how planets enter and leave the sub-Neptune bin over time. 

Recent studies have highlighted the importance of coupled orbital and interior evolution \citep{2009ApJ...702.1413M, 2024ApJ...972..159Y, 2025ApJ...979..218L, 2025ApJ...988..247S, 2026ApJ...997..138H, 2026ApJ...997..139H, 2026ApJ..1003...84I, 2026ApJ..1007L..38P, 2026arXiv260701315W}. Building on these efforts, we developed a unified framework called the ``coupled evolution model'' that self-consistently combines all three ingredients described above-- interior structure evolution, atmospheric mass-loss, and tidal migration. To our knowledge, these ingredients have not previously been combined in a single framework to conduct a population level study. 

Our focus is to test the hypotheses of \cite{2025AJ....169..208F} that atmospheric mass loss and tidal migration drive the evolution of sub-Neptune occurrence rates over time. 
HEM provides a compelling mechanism for bringing planets onto short-period orbits over timescales up to Gyrs. Recent work has shown that HEM can reproduce key demographic features, including the hot Neptune desert boundary and the newly identified ``Neptunian ridge" \citep{2024A&A...689A.250C, 2026A&A...709L..17C, 2026arXiv260620789Z}. Moreover, several close-in Neptunes exhibit residual eccentricities and large spin-orbit misalignments, including nearly polar orbits, suggestive of past dynamical excitation and tidal migration \citep[e.g.,][]{2021ApJ...916L...1A, 2023A&A...669A..63B, 2024A&A...690A.379K, 2024ApJ...972..159Y, 2025ApJ...979..218L, 2026arXiv260701315W}. Further supporting this picture, \citet{2025ApJ...988..247S} found stronger tidally induced radius inflation among misaligned Neptune-sized planets than aligned ones. Together, these results motivate HEM as a potentially important formation pathway for close-in Neptunes.

Thus, while other channels such as in-situ formation or disk-driven migration must also populate the close-in small-planet populations \citep[e.g.,][]{2013MNRAS.431.3444C,2021A&A...650A.152I}, 
we isolate HEM and ask what maximum contribution this formation channel can make to the age-dependent occurrence-rate evolution of close-in small planets. Because we intentionally select systems undergoing efficient HEM, our calculation provides an upper-limit estimate of the HEM contribution. This framing allows us to assess whether HEM can reproduce the observed age-dependent trends or whether modeling additional physics or formation channels may be required for understanding the current observations.

The paper is organized as follows. In \S~\ref{sec:model}, we describe the model including thermal evolution, atmospheric mass loss, and orbital evolution. In \S~\ref{sec:pop_synthesis}, we describe the construction of the synthetic planet population and the coupled evolution framework used to evolve this population. In \S~\ref{sec:results}, we present the main results and compare the model predictions with observations. In \S~\ref{sec:discussion}, we discuss occurrence-rate predictions for young super-Earths, the role of tidally induced radius inflation, and model caveats. We summarize our conclusions in \S~\ref{sec:conclusions}.

\section{Model} \label{sec:model}
Here we introduce the physical ingredients of our coupled evolution model. We first present the interior structure model, which determines how planetary radii evolve via thermal cooling in \S~\ref{sec:thermal_evol}. We then describe the photoevaporative mass-loss model in \S~\ref{subsec:massloss}, which governs atmospheric escape. Finally, we outline the orbital evolution model in \S~\ref{sec:orbital evol}. We introduce these processes separately before later coupling them together.
\subsection{Interior Structure and Thermal Evolution} \label{sec:thermal_evol}
We employ the method of \cite{2026ApJ...997..138H} to model the planetary interior structure and thermal evolution. Planets are assumed to consist of rocky, Earth-like cores 
surrounded by a H/He envelope with solar metallicity. Details of the microphysics employed in the structure models, (e.g. the equation of state and opacities) can be found in \cite{2026ApJ...997..138H}. 

We pre-compute a large grid of planet structures in thermal and mechanical equilibrium across four axes: core mass ($M_{\rm core}$), atmospheric mass fraction (${X_{\rm atm} \equiv M_{\rm atm}/M_{\rm core}}$), equilibrium temperature ($T_{\rm eq}$), and specific entropy ($S$) of the planet's innermost convective zone \citep[we do not account for non-convective interiors; e.g.][]{2022A&A...665A..12M}. Details of our structure grid boundaries are summarized in Table \ref{tab:structure_tables}. We note that the precise maxima and minima in our entropy axis depends on the other parameters (e.g. a low mass planet cannot retain an atmosphere with entropy $S{\sim}11$, while a massive planet can); the bounds recorded in Table \ref{tab:structure_tables} are approximate.

At each point in this four-dimensional grid, corresponding to a particular choice of ($M_{\rm core}, X_{\rm atm}, T_{\rm eq}, S$), the structure model provides the planet's radius, $R_p$, the cooling (outgoing) luminosity, $L$, set by conditions at the radiative convective boundary, and the thermal inertia term, $\int_{\rm conv}{Tdm}$. The specific entropy evolves according to
\begin{equation} \label{eq:dSdt}
    \frac{dS}{dt} = \frac{-L}{\int_{\rm conv}{Tdm}},
\end{equation}
where $dm$ is a differential mass element and the integral is taken over the convective portion of the envelope. The denominator represents the thermal inertia of the convective region and sets how rapidly the planet cools for a given luminosity. Thus, as the planet radiates energy, its interior entropy in the convective zone decreases and its radius contracts \citep[``following the adiabats"; e.g.][]{2006ApJ...650..394A}. We note that extra heating, e.g. tidal heating is not included here. We discuss tidally induced radius inflation later in \S~\ref{sec:tidal_heating_modeling}.

\begin{deluxetable}{CCCCCCc}\label{tab:structure_tables}
\caption{Parameters varied in planet structure calculations.}
\tablecolumns{4}
\tablewidth{0pt}
\tablehead{
\colhead{Parameter} &
\colhead{Value} & 
\colhead{Units} & 
}
\startdata
$M_{\rm core}$ & [2,20] & M_{\oplus} \\
\rm $X_{\rm atm}$ & [10^{-4},1] & ... \\
$T_{\rm eq}$ & $\{50,288,912,1600,2500\}$ & \rm K \\
$S$  & [4.9${-}$11] & k_{\rm B}/m_{\rm H}
\enddata
\end{deluxetable}

We construct an interpolation to the grid using a local linear k-nearest-neighbor (kNN) method. A KD-tree is used to identify the $k=8000$ nearest neighboring grid points \citep{1977ACM...KDTree}, and a distance-weighted local linear model is then fit to estimate $L$, $\int_{\rm conv}{Tdm}$, $R_p$ \citep{1988JASA...ClevelandDevlin}. Compared to direct high-dimensional interpolation, this scheme is faster for the repeated grid queries that will be required, does not require a strictly rectangular grid, and remains stable near gaps or irregular boundaries introduced by removing non-convective structures. Because the estimate is based on a local regression rather than a strict cell-based interpolant, it also permits modest extrapolation near the grid boundaries, which we use for small excursions outside the tabulated domain.
 

In our coupled evolution framework, the entropy evolution supplies the planet’s time-dependent radius, $R_p(S(t))$. At each timestep, we use the planet’s instantaneous properties to interpolate the precomputed structure grid and obtain $L$ and $\int_{\rm conv}{Tdm}$, which determine $dS/dt$ using Eq.~\ref{eq:dSdt}. We then update the entropy and use the new value of $S$, together with $M_{\rm core}, X_{\rm atm}, T_{\rm eq}$, to interpolate the grid and retrieve the corresponding radius, $R_p(t)$. Thus, by evolving $S(t)$, we obtain $R_p(t)$, which in turn affects both the atmospheric mass-loss rate and the dynamical evolution of the planet as described in \S\ref{subsec:massloss} \& \S\ref{sec:orbital evol}.

\subsection{Atmospheric Mass Loss}\label{subsec:massloss}
Mass loss is assumed to occur via photoevaporation, wherein high-energy X-ray and extreme ultraviolet (EUV) photons from the host star heat the planet's upper atmosphere, increasing the thermal energy of the gas and enabling it to escape the planet’s gravitational potential \citep{2003ApJ...598L.121L, 2017ApJ...847...29O}. We track the atmospheric mass fraction over time with \citep{2021MNRAS.503.1526R},
\begin{equation} \label{eq:dXdt}
    \frac{dX_{\rm atm}}{dt} = -\frac{X_{\rm atm}}{\tau_{\dot{X}}},
\end{equation}
where $\tau_{\dot{X}}$ is the characteristic atmospheric mass-loss timescale, $\tau_{\dot{X}} = X_{\rm atm}/\dot{X}_{\rm atm} = M_{\rm atm}/\dot{M}_{\rm atm}$. $\dot{M}_{\rm atm}$ is the atmospheric mass-loss rate due to photoevaporation. We consider a planet to be completely stripped once the $X_{\rm atm}$ value drops below $10^{-4}$, at which point its transit radius is effectively indistinguishable from that of a bare core.

For simplicity, we estimate $\dot{M}_{\rm atm}$ using the energy-limited approximation, which assumes that a fixed fraction of the incident stellar XUV flux is converted into work required to unbind atmospheric gas and is given by \citep{1981Icar...48..150W, 2004A&A...419L..13B, 2007A&A...472..329E},
\begin{equation}  \label{eq: Mdot}
    \dot{M}_{\rm atm} = \eta \frac{R_p^3L_{\rm XUV}}{4a^2GM_p},
\end{equation}
where $G$ is the gravitational constant, $L_{\rm XUV}$ is the high-energy luminosity of the host star, $M_p$ is the planet's mass, and $a$ is the semi-major axis. The parameter $\eta$ is the mass-loss efficiency, which we quantify by adopting a physically motivated scaling based on hydrodynamic simulations \citep{2012MNRAS.425.2931O}. Specifically, we parameterize $\eta$ as a function of the planetary escape velocity, $v_{\rm esc}$, as \citep{2017ApJ...847...29O, 2019AREPS..47...67O}
\begin{equation} \label{eq:eta}
    \eta = \eta_0\left(\frac{v_{\rm esc}}{25 \rm km s^{-1}} \right)^{-\alpha_{\eta}},
\end{equation}
where $\eta_0$ is a normalization constant and $\alpha_\eta$ is a power-law index. We adopt $\eta_0 = 0.17$, motivated by hydrodynamic simulations \citep[][see also \citealt{2019ApJ...874...91W}]{2012MNRAS.425.2931O,2014A&A...571A..94S}, and $\alpha_\eta = 0.5$ based on \citet{2021MNRAS.503.1526R}. In the standard formulation, a tidal correction factor $K_{\rm tide}$ is included to account for the reduction in the gravitational potential barrier due to tidal terms \citep[e.g.,][]{2007A&A...472..329E}. Here, we neglect this correction for simplicity, effectively assuming $K_{\rm tide} = 1$.

Observational \citep{1997ApJ...483..947G, 2005ApJ...622..680R, 2011ApJ...743...48W, 2012MNRAS.422.2024J, 2023ApJ...945..147N} and theoretical studies \citep{ 2015A&A...577L...3T, 2021A&A...649A..96J} have shown that the ratio $L_{\rm XUV}/L_{\rm bol}$, where $L_{\rm bol}$ is the total stellar (bolometric) luminosity, remains in a saturated regime for young stars and declines quickly as the star spins down. Motivated by this, we model the temporal evolution of the stellar high-energy luminosity as \citep{2017ApJ...847...29O},
\begin{equation} \label{eq:LXUV}
L_{\rm XUV}(t) =
\begin{cases}
L_{\rm sat}, & t < t_{\rm sat} \\
L_{\rm sat} \left(\dfrac{t}{t_{\rm sat}} \right)^{-(1 + \alpha_0)}, & t \ge t_{\rm sat}
\end{cases}
\end{equation}
where $\alpha_0 = 0.5$ and $t_{\rm sat} \sim 100~{\rm Myr}$. The saturation luminosity is given by
\begin{equation}
L_{\rm sat} \simeq 10^{-3.5} L_\odot \left( \frac{M_\star}{M_\odot} \right),
\end{equation}
where $M_\star$ is the mass of the host star. 

For planets on eccentric orbits, the instantaneous mass-loss rate varies along the orbit due to the changing stellar irradiation. To account for this, we approximate the orbit-averaged mass-loss rate as \citep{2002IJAsB...1...61W, 2016A&A...591A.106B}, 
\begin{equation} \label{eq:orbit_avg_mdot}
    \langle\dot{M}_{\rm atm}\rangle = \frac{\dot{M}_{\rm atm}}{\sqrt{1 - e^2}},
\end{equation}
where $e$ is the orbital eccentricity. This scaling captures the enhancement in the orbit-averaged XUV flux received by the planet and is used when coupling atmospheric escape to the orbital evolution. 

\subsection{Orbital Evolution} \label{sec:orbital evol}
Our goal is to test the hypothesis from \cite{2025AJ....169..208F} that HEM drives an increase in the sub-Neptune occurrence rate from young to intermediate ages. We approach this by isolating HEM and quantifying its maximum possible contribution. Planets in our synthetic population are excited onto highly eccentric orbits capable of undergoing HEM.
We emphasize that this is intentionally constructed to represent an efficient HEM limit. In reality, only a fraction of planets are expected to actually undergo HEM. The results in \S \ref{sec:results} should therefore be interpreted as an estimate of the maximum contribution that HEM channel can make to the age evolution of the close-in small-planet population, rather than as a prediction of the absolute occurrence rates.

HEM can be generated by several mechanisms, including planet--planet scattering \citep{2002Icar..156..570M, 2008ApJ...686..580C, 2012ApJ...751..119B} and secular perturbations from an outer companion. Here, we focus on von Zeipel--Lidov--Kozai (ZLK) oscillations, which operate in hierarchical three-body systems with large mutual inclinations and can drive coupled oscillations of eccentricity and inclination \citep{1910AN....183..345V,1962P&SS....9..719L,1962AJ.....67..591K}. We model the orbital evolution in two stages. First, planets undergo efficient ZLK oscillations, which are ultimately quenched by short-range forces (SRFs) including tides and general relativity, leaving the planet on a highly eccentric orbit immediately post-ZLK \citep{1997Natur.386..254H, 2002ApJ...576..894M, 2003ApJ...589..605W}. After this point, the planet decouples from the outer perturber and continues migrating while conserving angular momentum \citep{1981A&A....99..126H}. We discuss these two phases below.

\subsubsection{ZLK Oscillations} \label{sec:ZLK}

\begin{figure}
    \centering
    \includegraphics[width=\linewidth]{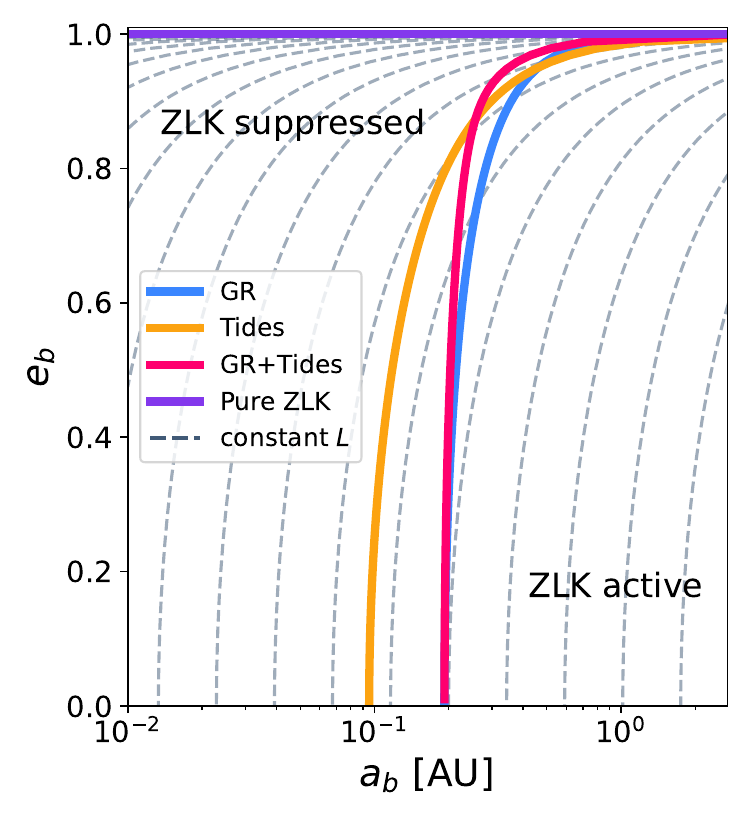}
    \caption{Maximum eccentricity reached during ZLK excitation as a function of the inner planet's semi-major axis for an example system. The purple curve shows the maximum eccentricity reached in the absence of SRFs. The blue and orange curves show the SRF-limited maximum eccentricity when only GR precession or tidal precession is included, respectively. The pink curve includes both GR and tidal precession and defines the SRF-limited post-quenching eccentricity. Regions to the left of these curves are effectively ZLK-suppressed and the subsequent tidal migration proceeds approximately along a constant-angular-momentum track, shown by the gray dashed curves.}
    \label{fig:a_e_contours}
\end{figure}

Here we briefly summarize the aspects of ZLK dynamics relevant for our analysis (see \citealt{2016ARA&A..54..441N} for a review). We assume the existence of an outer perturber (hereafter body c), creating a hierarchical three-body system. The secular hierarchical approximation is valid if the system satisfies the empirical dynamical stability criterion given by \citet{2001MNRAS.321..398M}
\begin{equation} \label{eq:stability}
\begin{aligned}
\frac{a_c}{a_b} >\;&
2.8\left(1+\frac{M_c}{M_\star+M_p}\right)^{2/5}
\left(1+e_c\right)^{2/5}
\left(1-e_c\right)^{-6/5} \\
&\times
\left[1-0.3\left(\frac{\psi_{\rm bc}}{180^\circ}\right)\right],
\end{aligned}
\end{equation}
where $M_c$ is the mass of body c, $a_c$ and $a_b$ are the semi-major axes of body c and the planet, respectively, $e_c$ is eccentricity of body c, and $\psi_{bc}$ is the mutual inclination between the inner and outer orbits. 
When $\psi_{bc}$ lies within a critical range ($39.2^\circ \le \psi_{bc} \le 140.8^\circ$ for the classic quadrupolar, inner test-particle case), the inner orbit undergoes oscillations in eccentricity and inclination \citep{2000ApJ...535..385F, 2016MNRAS.460.1086M}. 


If the influence of SRFs such as general relativity and tidal precession is negligible, the maximum eccentricity achieved by the inner planet during ZLK oscillations ($e_{b, \max}$), assuming an initially circular orbit, depends only on $\psi_{bc}$ and is given by $e_{b, \max} = \sqrt{1-(5/3)\cos^2\psi_{bc}}$ \citep{2007ApJ...669.1298F}. However, in the presence of SRFs, the growth of eccentricity is suppressed, and thus the maximum eccentricity achieved is reduced. In the quadrupole limit, $e_{b, \max}$ can be obtained analytically by solving the equation \citep{2000ApJ...535..385F, 2015MNRAS.447..747L}:
\begin{equation} \label{eq:ebmax}
\begin{aligned}
\epsilon_{\rm GR}\left(\frac{1}{j_{b,\min}} - 1\right)
&+ \frac{\epsilon_{\rm tide}}{15}
\left[
\frac{1 + 3e_{b,\max}^2 + \frac{3}{8}e_{b,\max}^4}{j_{b,\min}^9}
- 1
\right] \\
&=\; \frac{9}{8}\frac{e_{b,\max}^2}{j_{b,\min}^2}
\left(
j_{b,\min}^2 - \frac{5}{3}\cos^2 \psi_{bc}
\right),
\end{aligned}
\end{equation}
where $j_{b,\min} = \sqrt{1 - e_{b,\max}^2}$. We neglect the contribution from rotation-induced planetary oblateness here, as it is typically at least an order of magnitude smaller than the dominant GR and tidal bulge precession terms. The dimensionless parameters $\epsilon_{\rm GR}$ and $\epsilon_{\rm tide}$ are related to apsidal precession due to GR and the tidal bulge raised on the planet by the host star, respectively. These are given by \citep{2001ApJ...562.1012E}:
\begin{subequations}
\begin{align}
\epsilon_{\rm GR} &= \frac{3G(M_\star + M_p)^2a_c^3(1-e_c^2)^{(3/2)}}{a_b^4c^2M_c},\\
\epsilon_{\rm tide} &= \frac{15M_\star(M_\star + M_p)a_c^3(1-e_c^2)^{(3/2)}k_{2,p}R_p^5}{a_b^3M_p M_c},
\end{align}
\end{subequations}
where $c$ is the speed of light and $k_{2, p}$ is the planet's Love number.  

When the high eccentricity episodes drive the inner planet close enough to the star, they drive tidal migration. The migration timescale can be estimated by using the fraction of time spent near eccentricity peaks, $\sim \sqrt{1 - e_{b, \max}^2}$, yielding the orbit-averaged decay rate \citep{2019MNRAS.484.5645V}
\begin{equation} \label{eq:t_ZLK}
        t_{\rm ZLK}^{-1} = \left(|t_{\rm ST}^{-1}|\sqrt{1 - e^2} \right)_{\rm e_{b, \max}}
\end{equation}
where $t_{\rm ST}^{-1}$ is the timescale of orbital decay due to static tides and is given by \citep{1973Ap&SS..23..459A, 1981A&A....99..126H},
\begin{equation} \label{eq:decay_timescale}
    \begin{aligned}
        t_{\rm ST}^{-1} = \frac{9n^2}{Q'_p}\frac{M_\star}{M_p}\left(\frac{R_p}{a}\right)^5\left[N_a(e) - \frac{N^2(e)}{\Omega(e)} \right],
    \end{aligned}
\end{equation}
where $Q'_p$ is the reduced tidal quality factor of the planet, $n$ is the planet's mean motion, and 
\begin{equation} \label{n_a}
    N_{\rm a}(e) = \frac{1 + \frac{31}{2}e^2 + \frac{255}{8}e^4 + \frac{185}{16}e^6 + \frac{25}{64}e^8}{(1-e^2)^{\frac{15}{2}}},
\end{equation}
\begin{equation} \label{n(e)}
    N(e) = \frac{1 + \frac{15}{2}e^2 + \frac{45}{8}e^4 + \frac{5}{16}e^6}{(1-e^2)^6},
\end{equation}
\begin{equation} \label{eq:omega(e)}
    \Omega(e) = \frac{1 + 3e^2 + \frac{3}{8}e^4}{(1 - e^2)^{\frac{9}{2}}}.
\end{equation}


ZLK oscillations are quenched once the apsidal precession induced by SRFs become comparable to, or exceeds, the precession rate driven by perturbations from the outer companion. Figure~\ref{fig:a_e_contours}, based on a similar figure in \citet{2025ApJ...979..218L}, illustrates this behavior for an example system with $M_p = 4 M_\oplus$, $M_c = 0.3~M_\odot$, $M_\star = 1~M_\odot$, $e_c = 0.85$, $a_c = 100$~au, $X_{\rm atm} = 0.04$, $\psi_{bc} = 90^\circ$, and $k_{2,p} = 0.25$. The full SRF-limited eccentricity is obtained by solving for $e_{b, \max}$ in Eq.~\ref{eq:ebmax} that includes both GR and tidal precession simultaneously, and is shown by the pink curve. The pink curve defines the orbital state of the planet post-ZLK quenching, from which the planet begins subsequent tidal migration approximately along a constant-angular-momentum track, shown by the gray dashed curves.  


\subsubsection{Tidal Migration} \label{sec:tidal_migration}
After ZLK quenching, we neglect further gravitational perturbations from body c and evolve the planet as a two-body star--planet system. Each planet is initialized at its post-ZLK orbital state, $(a_0,e_0)$, where $e_0$ is the SRF-limited maximum eccentricity obtained from Eq.~\ref{eq:ebmax} at the quenching semi-major axis $a_0$. The subsequent orbital evolution is governed by equilibrium tides in the constant time-lag approximation \citep{1980A&A....92..167H,2010A&A...516A..64L}.

We include only tides raised by the star on the planet and neglect tides raised by the planet on the star, since for Neptune-sized planets the latter contribution is typically smaller by a factor $\lesssim 10^{-4}$ and therefore negligible \citep{2025ApJ...988..247S}. For simplicity, we also assume zero planetary obliquity. Under these assumptions, the semi-major axis and eccentricity evolve as follows \citep{2010A&A...516A..64L}:
\begin{align} \label{da_dt}
       \frac{1}{a}\frac{da}{dt} = \frac{4a}{GM_\star M_p} K\left [N(e)\frac{\omega_p}{n} - N_a(e) \right],
\end{align}
\begin{align} \label{de_dt}
       \frac{1}{e}\frac{de}{dt} = \frac{11a}{GM_\star M_p}K\left [\Omega_e(e)\frac{\omega_p}{n} - \frac{18}{11}N_e(e) \right],
\end{align}
where $N_a(e)$, and $N(e)$ are given in Eqs.~\ref{n_a} \& \ref{n(e)}. $K$ is the characteristic luminosity scale given by
\begin{equation} \label{eq: K}
    K = \frac{9n}{4Q'_p}\left(\frac{GM_{\star}^2}{R_p}\right) \left(\frac{R_p}{a}\right)^6, 
\end{equation}
and $\Omega_e(e), N_e(e)$ are defined as:
\begin{equation}
    \Omega_e(e) = \frac{1 + \frac{3}{2}e^2 + \frac{1}{8}e^4}{(1 - e^2)^5},
\end{equation}
\begin{equation} 
    N_e(e) = \frac{1 + \frac{15}{4}e^2 + \frac{15}{8}e^4 + \frac{5}{64}e^6}{(1-e^2)^{13/2}}.
\end{equation}
We also assume that the planet's spin frequency, $\omega_p$ has reached an equilibrium state such that $d\omega/dt = 0$ and is given by $\omega_p = \omega_{\rm eq} = nN(e)/\Omega(e)$, where $\Omega(e)$ is defined in Eq. \ref{eq:omega(e)}.

We note that our coupled evolution model does not model the ZLK phase explicitly; rather, we use the equations from \S\ref{sec:ZLK} to initialize planets in a post-ZLK state and then migrate them according to the equations just presented. Further details are given in the next section. 

\section{Population Synthesis} \label{sec:pop_synthesis}
The construction of our synthetic planet population proceeds in six steps summarized below. 
\begin{itemize}
    \item \textbf{Step 1: Sample the star and planet properties.}
    We draw the $\{M_\star, P_{\rm orb}, M_{\rm core}, X_{\rm atm}^{\rm init}, Q'_p\}$ from distributions specified later in the section.
    \item \textbf{Step 2: Sample the perturber parameters.}
    For each system, we sample the parameters of body c, $\{M_c,P_c,e_c\}$. We also draw $\psi_{bc}$, restricting it to values for which efficient ZLK excitation can occur.
    \item \textbf{Step 3: Select dynamically stable systems.}
    Using the sampled parameters, we retain only systems that satisfy the hierarchical stability criterion (Eq. \ref{eq:stability}).
    \item \textbf{Step 4: Determine post-ZLK orbital state.}
    Assuming ZLK oscillations are quenched at the semi-major axis corresponding 
    to the sampled inner orbital period ($P_{\rm orb}$), we solve for the SRF-limited maximum eccentricity using Eq.~\ref{eq:ebmax}. The resulting pair, $(a_0,e_0)$, is defined as the post-ZLK orbital state.
    \item \textbf{Step 5: Apply physical and timescale cuts.}
    We remove planets with pericenter distances inside the Roche limit and/or with ZLK-driven migration timescales longer than $100$~Myr.
    \item \textbf{Step 6: Simulate coupled evolution for surviving population.}
    The remaining systems comprise the post-ZLK population, and we model their coupled thermal, atmospheric mass-loss, and tidal evolution.
\end{itemize}
Each of these steps are described in detail in \S\ref{sec:properties}--\S\ref{sec:coupled evolution}.

\begin{table*}
\centering
\caption{Summary of the sampled parameters used for population synthesis}
\label{tab:pop_synthesis_distributions}
\begin{tabular}{lll}
\hline
\textbf{Quantity} & \textbf{Distribution / value} & \textbf{Notes} \\
\hline
$M_\star$ & $\mathcal{N}(1.04\,M_\odot,\,0.15^2\,M_\odot^2)$ 
& Host-star mass; CKS-motivated distribution \\

$P_{\rm orb}$ & Smooth broken power law 
& $P_0=5.75$ d, $k_1=2.31$, $k_2=-0.08$; restricted to $[2,1000]$ days \\

$M_{\rm core}$ & Log-normal 
& $\mu_M=3.72\,M_\oplus$, $\sigma_M=0.44$; restricted to $[2,20]\,M_\oplus$ \\

$X_{\rm atm}^{\rm init}$ & Log-normal 
& $\mu_X=0.04$, $\sigma_X=0.51$ \\

$Q'_p$ & Discrete uniform 
& $Q'_p \in \{10^4,10^5,10^6\}$ \\

Core composition & Fixed 
& Earth-like rock/iron core composition \\

\hline
$q$ & $\mathcal{N}(0.23,\,0.42^2)$ 
& Companion-to-inner-system mass ratio; reject $q\leq 0$ \\

$M_c$ & $M_c=q(M_p+M_\star)$ 
& Companion mass derived from sampled $q$ \\

$P_c$ & Log-normal 
& $\mu_{\log_{10}P_c}=5.03$, $\sigma_{\log_{10}P_c}=2.28$ \\

$e_c$ & Rayleigh or thermal 
& Rayleigh for $P_c<1000$ d; thermal distribution for $P_c>1000$ d \\

$\psi_{bc}$ & Truncated normal 
& Centered at $90^\circ$ with $\sigma=5^\circ$, restricted to $39.2^\circ$--$140.8^\circ$ \\
\hline \\
\end{tabular}
\end{table*}

\subsection{Star and planet properties}
\label{sec:properties}
We first sample the intrinsic properties of the host star and inner planet. For each system, we draw $M_\star$, $P_{\rm orb}$, $M_{\rm core}$, $X_{\rm atm}^{\rm init}$, and $Q'_p$.

For the stellar mass, we adopt a Gaussian distribution motivated by the California-Kepler Survey (CKS) \citep{2017AJ....154..109F},  
$M_\star \sim \mathcal{N}(\mu_{M_\star}, \sigma_{M_\star}^2)$, where \(\mu_{M_\star} = 1.04\,M_\odot\) and \(\sigma_{M_\star} = 0.15\,M_\odot\), following \citet{2021MNRAS.503.1526R}. 

$P_{\rm orb}$ is sampled from the long-period branch of a broken power-law distribution motivated by occurrence rate observations \citep{2012ApJS..201...15H, 2013ApJ...766...81F, 2015ApJ...807...45D, 2018AJ....155...89P}. 
We restrict the range to $P_{\rm orb}\in[50,1000]$ days and sample from
\begin{equation}
\frac{dN}{dP_{\rm orb}} \propto
\left(\frac{P_{\rm orb}}{P_0}\right)^{k},
\end{equation}
with $k=-0.08$. 
Because this distribution is motivated by mature planet populations, it serves as an observationally motivated prior for the reservoir of planets available to undergo ZLK-driven HEM. The true primordial distribution of HEM progenitors is poorly constrained. 

The planet's core mass and initial atmospheric mass fraction are drawn from log-normal distributions adopted by \citet{2019ApJ...874...91W}; similar functional forms were also used by \citet{2021MNRAS.503.1526R}. These distributions are given by:

\begin{equation}
\frac{dN}{d\log M_{\rm core}}
\propto
\exp\!\left[
-\frac{(\log M_{\rm core}-\log \mu_M)^2}{2\sigma_M^2}
\right],
\end{equation}
\begin{equation}
\frac{dN}{d\log X_{\rm atm}^{\rm init}}
\propto
\exp\!\left[
-\frac{(\log X_{\rm atm}^{\rm init}-\log \mu_X)^2}{2\sigma_X^2}
\right],
\end{equation}
where $\mu_M = 3.72\,M_\oplus$, $\sigma_M = 0.44$, and
$\mu_X = 0.04$, $\sigma_X = 0.51$. We further restrict the sampled core masses to $M_{\rm core} \in [2, 20] M_\oplus$, corresponding to the range most relevant for planets expected to evolve into sub-Neptune population over $\sim$Gyr timescales. 

Typical values for the reduced tidal quality factor (\(Q'_p\)) are highly uncertain. For this analysis, we draw \(Q'_p\) uniformly from the discrete set \(\{10^4,\,10^5,\,10^6\}\), based on Solar System planets \citep{1999ssd..book.....M, 1990Icar...85..394T, 2008Icar..193..267Z}, and consistent with recent population-level studies that model sub-Neptune and Neptune-size planets with $Q'_p$ values in a similar range \citep{2018AJ....155..157P, 2020ApJ...897....7M, 2025ApJ...988..247S}.  

We adopt an Earth-like rock/iron core composition, following previous evolutionary studies of sub-Neptunes \citep{2017ApJ...847...29O, 2017AJ....154..109F}. We note, however, that the true sub-Neptune population may exhibit a broader diversity of interior compositions \citep[e.g.,][]{2021JGRE..12606639B, 2022Sci...377.1211L, 2024A&A...688A..59P}. 

\subsection{Perturber Parameters}
\begin{figure}
    \centering
    \includegraphics[width=\linewidth]{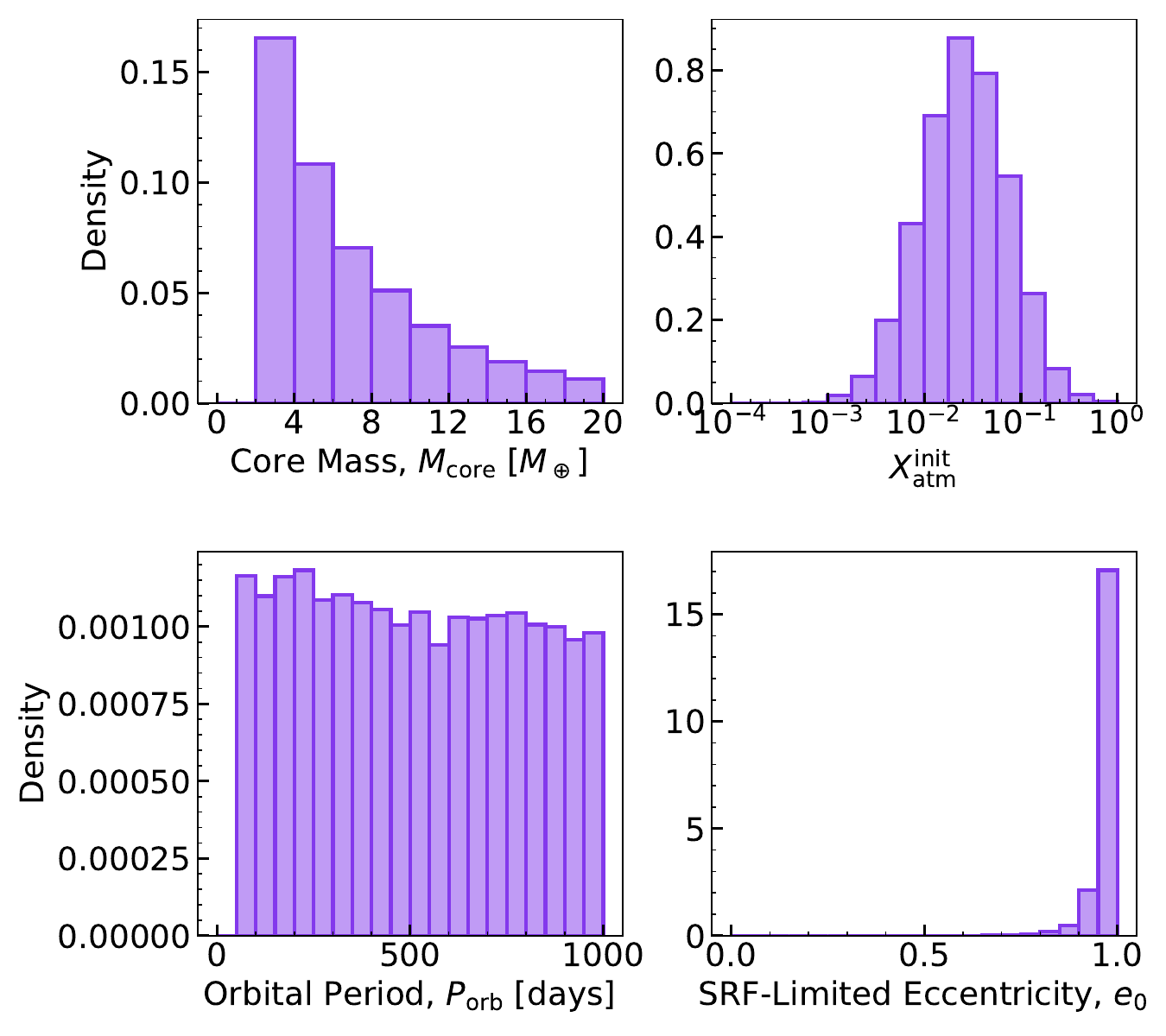}
    \caption{Distributions of the sampled planet properties and post-ZLK quenching orbital conditions for the planet population that we evolve. The top panels show the core mass, $M_{\rm core}$, and initial atmospheric mass, fraction, $X_{\rm atm}^{\rm init}$, while the bottom panels show the orbital period, $P_{\rm orb}$, and SRF-limited eccentricity at quenching, $e_0$. 
The concentration of systems at high $e_0$ reflects the selection of planets undergoing efficient ZLK-driven high-eccentricity migration.}
    \label{fig:planet_distributions}
\end{figure} 
We next assign an outer companion, body c, to each star–planet system. This companion provides the external perturbation to drive ZLK oscillations. The parameters $P_c$, $M_c$, $e_c$, and $\psi_{bc}$ determine whether the system can undergo efficient ZLK excitation.

\(M_c\) is sampled by drawing a mass ratio \(q\) from a normal distribution following \citet{2007ApJ...669.1298F}, $q \sim \mathcal{N}(\mu_q = 0.23,\, \sigma_q = 0.42)$. We reject unphysical draws with \(q \leq 0\). The companion mass is then computed as $M_c = q\,(M_p + M_\star)$, where $M_p = M_{\rm core}(1 + X_{\rm atm})$ is defined from the sampled $M_{\rm core}$ and $X_{\rm atm}^{\rm init}$. 

We draw the outer-companion orbital period, $P_c$ (in days) from a log-normal distribution with $\mu_{\log_{10} P_c}=5.03$ and $\sigma_{\log_{10} P_c} = 2.28$, following \citet{2010ApJS..190....1R}, who characterized the period distribution of solar-type stellar companions using combined spectroscopic and imaging surveys. 

The distribution for \(e_c\) depends on $P_c$ \citep{1991A&A...248..485D}. For systems with \(P_c < 1000\) days, $e_c$ is drawn from a Rayleigh distribution, $p(e)de = (e/\sigma_c^2) \exp{(-e^2/2\sigma_c^2)}de$,
with \(\langle e^2 \rangle^{1/2} = 0.33\), corresponding to \(\sigma_e = \langle e^2 \rangle^{1/2}/\sqrt{2}\). For wider systems with \(P_c > 1000\,\mathrm{d}\), eccentricities are drawn from an Ambartsumian (thermal) distribution, $p(e)\,de = 2e\,de$ \citep{2007ApJ...669.1298F}.

$\psi_{bc}$ is drawn from a truncated normal distribution centered at \(90^\circ\) with a standard deviation of \(5^\circ\), restricted to the classical ZLK window of \(39.2^\circ \leq \psi_{bc} \leq 140.8^\circ\). This preferentially samples configurations that efficiently excite high eccentricities. 

A summary of sampling distributions for the star, planet, and body c are presented in Table~\ref{tab:pop_synthesis_distributions}.

\subsubsection{Dynamical stability and post-ZLK orbital state}
\begin{figure}
    \centering
    \includegraphics[width=\linewidth]{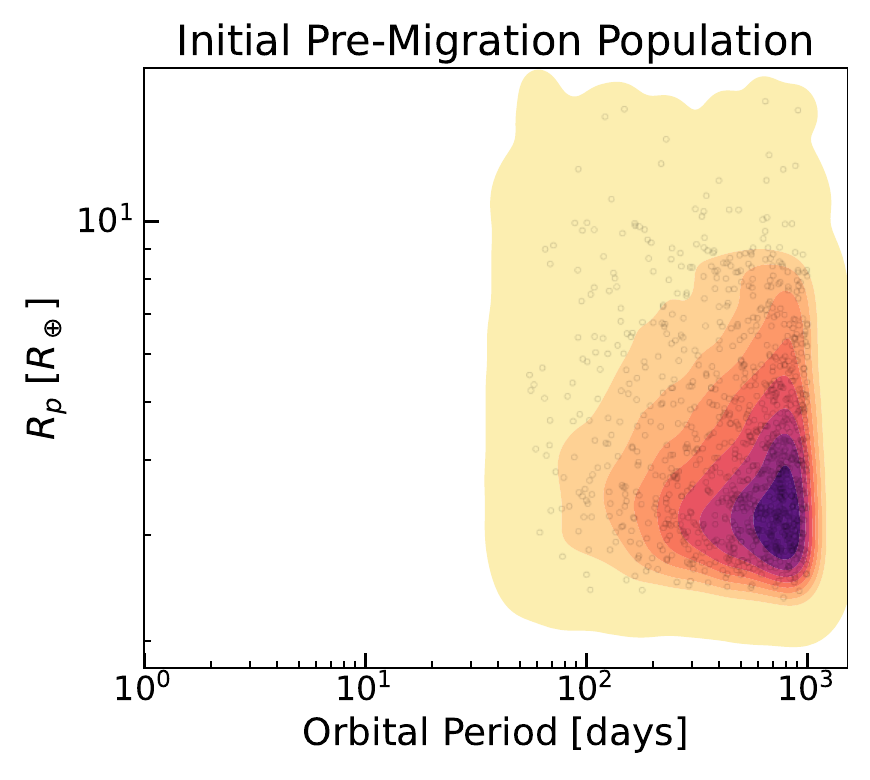}
    \caption{Initial period-radius distribution of the planet population before coupled evolution. The shaded background shows a Gaussian KDE of the full modeled sample, while gray open circles show a random subset of $\sim 2000$ planets for visual reference. Each planet is evaluated at the start of its coupled thermal, atmospheric, and tidal evolution, so the ensemble combines systems with different ZLK-quenching/onset times rather than representing a single fixed-age snapshot.}
    \label{fig:pre migration pop}
\end{figure}
After sampling the system parameters, we require the resulting hierarchical triple to satisfy the dynamical stability criterion given by Eq.~\ref{eq:stability}. Systems that fail this condition are rejected and resampled. We then determine the orbital state of the inner planet at ZLK quenching. 
We assume that ZLK oscillations are quenched at the semi-major axis corresponding to the sampled $P_{\rm orb}$, and we calculate the SRF-limited maximum eccentricity by solving for $e_{b, \max}$ in Eq.~\ref{eq:ebmax}. 
The semi-major axis and eccentricity pair, $(a_0, e_0)$, is the orbital state at ZLK quenching and is used as the initial orbital condition for subsequent evolution. 

\subsection{Physical and Timescale Cuts} \label{subsec:cuts}
Lastly, we apply two additional cuts to isolate systems that can plausibly contribute to the sub-Neptune population through efficient ZLK-driven migration. First, we remove systems whose pericenter distance lies inside the Roche limit, $a_{\rm roche} > a_0(1-e_0)$, where $a_{\rm roche} = 2.44R_p(M_\star/M_p)^{1/3}$, since these planets are expected to undergo tidal disruption. Second, we remove systems with $t_{\rm ZLK} > 100~$Myr. This cut isolates planets for which ZLK-driven migration occurs early. 
We adopt $t_{\rm ZLK}$ as an estimate for the ZLK quenching time and use it as an initial time for subsequent coupled thermal, atmospheric, and tidal evolution.

From an initial sample of $\sim 200,000$ systems, we retain $\sim 20,000$ viable systems that satisfy both criteria. We emphasize that this filtered sample should not be interpreted as representative of all planetary systems. Rather, it represents the subset dynamically capable of undergoing efficient ZLK-driven HEM while avoiding tidal disruption, allowing our model to provide an upper-limit on the contribution of HEM channel to sub-Neptune occurrence rates.


Following all the cuts, the sampled distributions of $M_{\rm core}$, $X_{\rm atm}^{\rm init}$, $P_{\rm orb}$, and $e_0$, are shown in Figure~\ref{fig:planet_distributions}.
\subsection{Coupled Evolution} \label{sec:coupled evolution}
\begin{figure}
    \centering
    \includegraphics[width=\linewidth]{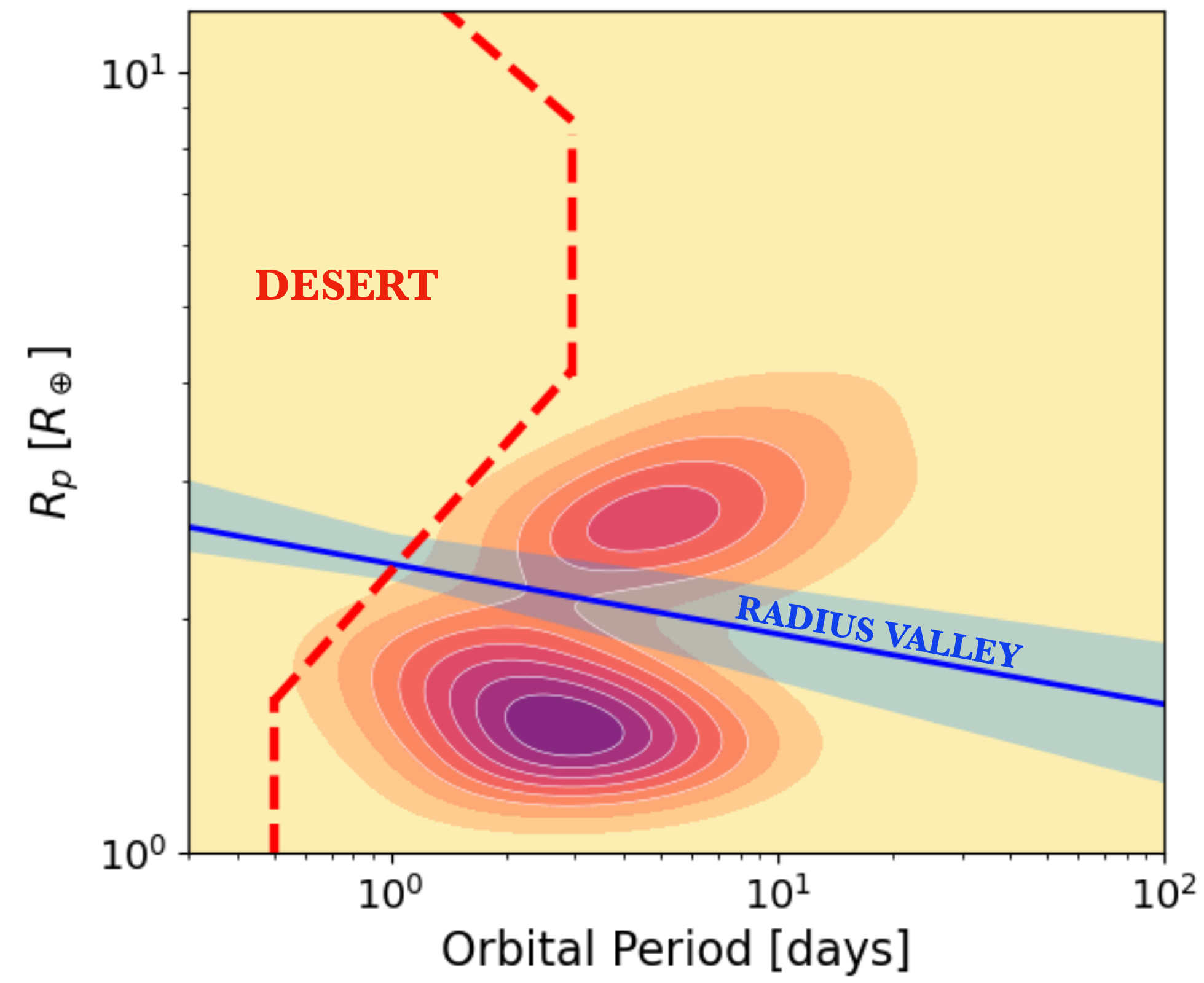}
    \caption{Final period-radius distribution of the evolved synthetic population at 3 Gyr. The contours show the Gaussian KDE of the model population. The blue line and shaded band mark the observed radius valley relation and its uncertainty from \citet{2018MNRAS.479.4786V}, while the red dashed curve shows the approximate boundary of the hot-Neptune desert from \citet{2024A&A...689A.250C}. 
    }
    \label{fig:demographic_features}
\end{figure}
We evolve each system in our synthetic population forward to 3 Gyr. The starting time depends on the system as $t_{\rm start} = 10~{\rm Myr} + t_{\rm ZLK}$, where $10~{\rm Myr}$ represents a fiducial disk lifetime (which we do not model). 
The second term, $t_{\rm ZLK}$, is our estimate of the time required for ZLK-driven excitation to be quenched by SRFs obtained from Eq.~\ref{eq:t_ZLK}. Since $t_{\rm ZLK}$ is estimated separately for each system, different planets begin their coupled evolution at different absolute ages. At $t_{\rm start}$, each system is initialized with parameters $M_\star$, $M_{\rm core}$, $X_{\rm atm}^{\rm init}$, post-quenching orbital elements $(a_0, e_0)$, and a high initial entropy, $S = 8$ $k_{\rm B}/m_{\rm H}$, appropriate for young planets. We do not account for very early atmospheric loss or ``boil-off,'' which could produce lower initial entropies \citep{2016ApJ...817..107O}. The initial period-radius distribution of the progenitor population is shown in Figure~\ref{fig:pre migration pop}. 
\begin{figure} 
    \centering
    \includegraphics[width = \linewidth]{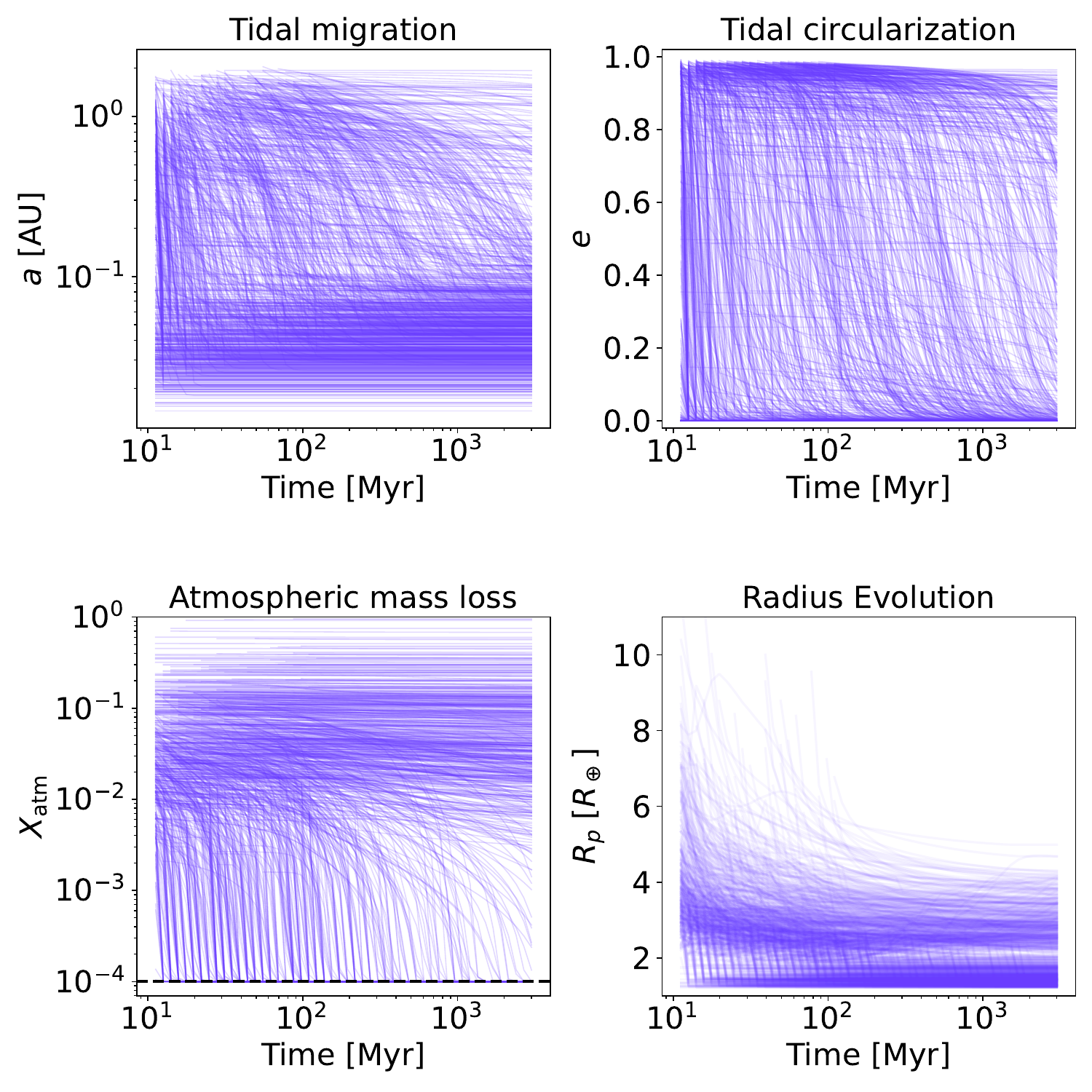}
    \caption{Coupled evolutionary tracks for 1000 randomly selected systems from the synthetic population. The four panels show semi-major axis evolution, eccentricity damping, atmospheric mass loss, and radius evolution. The dashed line at $X_{\rm atm} = 10^{-4}$ in the atmospheric mass loss evolution panel, marks the stripping threshold, below which planets are treated as bare rocky cores.}
    \label{fig:evolution_paths}
\end{figure}

We then evolve the planets using our coupled model for thermal contraction, atmospheric mass loss, and tidal orbital evolution. We solve four coupled ordinary differential equations: the entropy evolution equation, $dS/dt$ (Eq.~\ref{eq:dSdt}); the atmospheric mass-loss equation, $dX_{\rm atm}/dt$ (Eq.~\ref{eq:dXdt}), using the orbit-averaged mass-loss rate from Eq.~\ref{eq:orbit_avg_mdot}; and the equilibrium-tide equations for $da/dt$ (Eq.~\ref{da_dt}) and $de/dt$ (Eq.~\ref{de_dt}).
\begin{figure*}
    \centering
    \includegraphics[width=\linewidth]{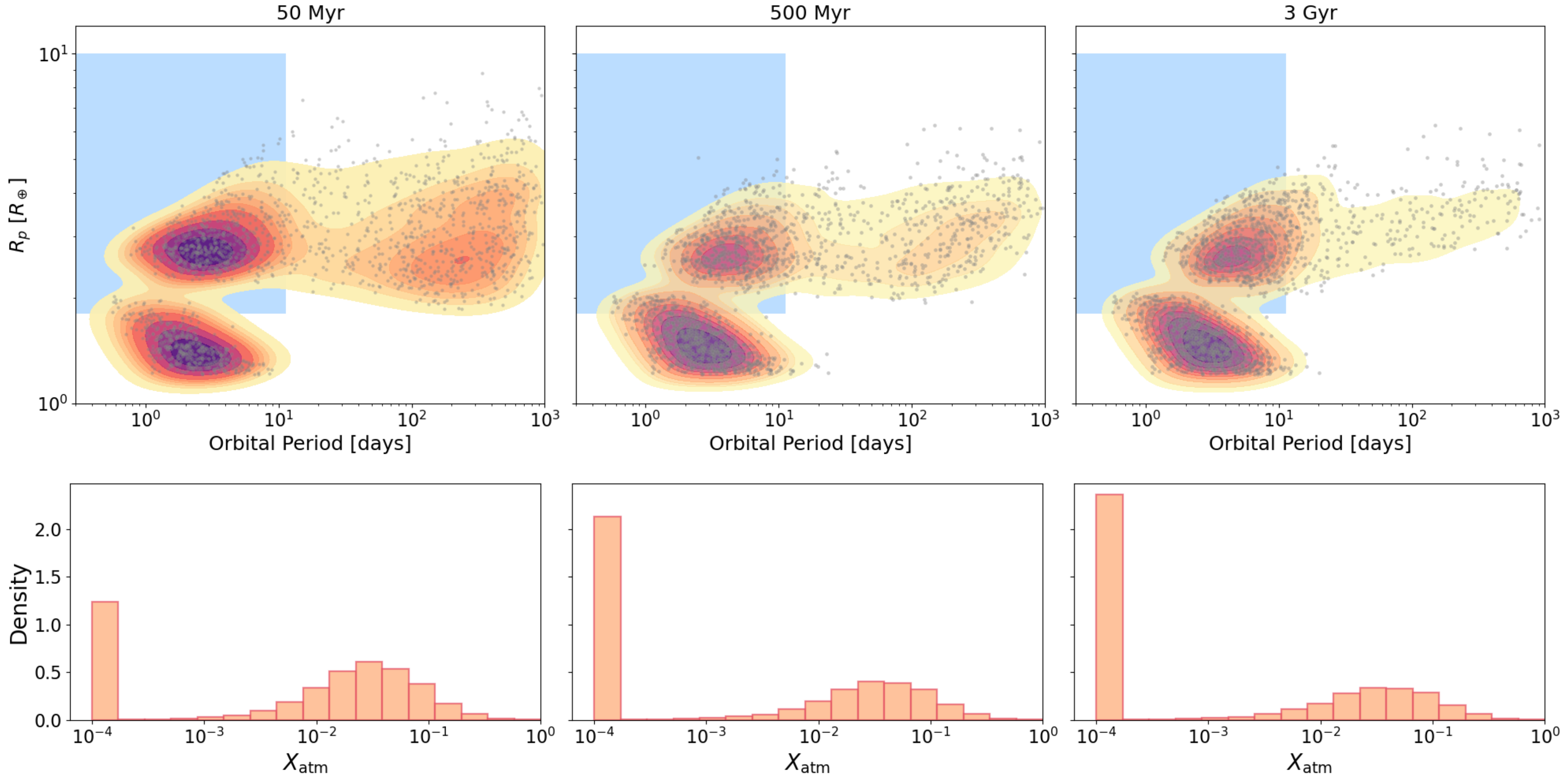}
    \caption{Snapshots of the synthetic planet population at three ages: 50 Myr, 500 Myr, and 3 Gyr. The upper panels show Gaussian KDEs of the population in the $P_{\rm orb}-R_p$ plane, with $\sim 2000$ randomly selected planets overplotted as grey points for visual reference. The blue shaded region marks the sub-Neptune bin used to compute occurrence rates, defined by $P_{\rm orb} < 12$ days and $1.8 < R_p/R_\oplus < 10$. The lower panels show the corresponding atmospheric mass fractions, highlighting the growth of the stripped-cores that populate the super-Earth regime. The pile-up at $X_{\rm atm} = 10^{-4}$ reflects our numerical stripping floor, below which planets are treated as fully stripped rocky cores.}
    \label{fig:3snpashots}
\end{figure*}
At each timestep, we self-consistently update the planet’s radius, total mass, and equilibrium temperature. The evolved variables are $S(t)$, $X_{\rm atm}(t)$, $a(t),$ and $e(t)$. $M_{\rm core}$ is held fixed throughout the integration. The equilibrium temperature is updated from the instantaneous orbital elements as it is required for thermal evolution. For eccentric orbits, we use the orbit-averaged $T_{\rm eq}$ value \citep{2022RNAAS...6...56Q}, 
\begin{equation}
    T_{\rm eq}(t) = T_\star\sqrt{\frac{R_\star}{2a(t)}}[1 - e^2(t)]^{-1/8},
\end{equation}
where we assume full heat redistribution and a fixed bond albedo $A_B = 0$. Nonzero albedos would reduce $T_{\rm eq}$ by a factor of $(1 - A_B)^{1/4}$. 
We estimate stellar radii ($R_\star$) and effective temperatures ($T_\star$) using approximate scalings, $R_\star \approx R_\odot(M_\star/M_\odot)^{0.8}$, and $T_\star \approx T_\odot(M_\star/M_\odot)^{0.48}$ from empirical mass-radius and mass-temperature relations for main-sequence stars \citep{1990sse..book.....K, 1996ima..book.....C}. We then compute the planet radius by interpolating the precomputed structure grid, $R_p(t) = R_p[M_{\rm core}, X_{\rm atm}(t), T_{\rm eq}(t), S(t)]$, as described in Section~\ref{sec:thermal_evol}. The total planet mass is updated separately using the instantaneous atmospheric mass fraction, $M_p(t) = M_{\rm core}[1+X_{\rm atm}(t)]$. These updated values of $R_p(t), M_p(t),$ and $T_{\rm eq}(t)$ are then used in the atmospheric mass loss (Eq.~\ref{eq: Mdot}) and equilibrium tide equations (Eq.~\ref{da_dt} \& Eq.~\ref{de_dt}), ensuring that the thermal, atmospheric, and orbital evolution remain coupled throughout the integration. 

When $X_{\rm atm} < 10^{-4}$, we classify the planet as fully stripped and set $dX_{\rm atm}/dt = 0$ for the remainder of the integration. From that point onward, the planet is treated as a bare Earth-composition core, and its radius is computed from a solid planet mass-radius relation from \citet{2007ApJ...659.1661F} with iron mass fraction, $X_{\rm iron} = 0.33$ and ice mass fraction, $X_{\rm ice} = 0$. Example evolutionary tracks for $\sim 1000$ randomly selected systems from our synthetic population are shown in Figure~\ref{fig:evolution_paths}.

\section{Results} \label{sec:results}

\subsection{Demographic Features of the Evolved Population}
The synthetic population of ${\sim20,000}$ systems is evolved from
$t_{\rm start}$ to 3 Gyr using the coupled evolution framework described in \S~\ref{sec:coupled evolution}. Before measuring the age-dependent occurrence rate, we first examine the final period-radius distribution at $\sim 3~{\rm Gyr}$ to check the overall behavior of the model. As shown in Figure~\ref{fig:demographic_features}, the evolved population qualitatively recovers two key demographic features, the radius valley \citep{2017ApJ...847...29O, 2017AJ....154..109F} and the hot-Neptune desert \citep{2011ApJ...727L..44S, 2016A&A...589A..75M}. In particular, we find a dearth of planets near $\sim 2R_\oplus$, broadly consistent with the observed radius valley \citep[e.g.,][]{2018MNRAS.479.4786V}, separating super-Earths that have lost most of their primordial H/He envelopes from sub-Neptunes that retain non-negligible atmospheres. The population also remains sparse in the hot-Neptune desert region (approximate $P-R_p$ boundaries following \citealt{2024A&A...689A.250C}), suggesting that the coupled effects of atmospheric escape, thermal contraction, and tidal evolution limit the survival of highly irradiated Neptune-size planets at short orbital periods. The recovery of these broad demographic features is encouraging and further motivates using the coupled evolution framework for modeling the sub-Neptune occurrence rate evolution with age.

\subsection{Time Evolution of the Synthetic Population} \label{subsec:timeevol}
To examine the time-dependent evolution, we first consider snapshots at 50~Myr, 500~Myr, and 3~Gyr, as shown in Figure~\ref{fig:3snpashots}. The blue shaded region marks our close-in sub-Neptune bin. By 50~Myr, $\sim 32\%$ of the  systems reach short orbital periods, and the population already begins to separate into sub-Neptunes that retain H/He envelopes and stripped super-Earths. This early atmospheric stripping is also evident from the pile-up near $X_{\rm atm}\simeq 10^{-4}$ in the lower panel of Figure~\ref{fig:3snpashots}.

The subsequent evolution is governed by two competing effects: (i) tidal migration, which moves additional planets into the sub-Neptune bin, and (ii) atmospheric mass loss \& thermal contraction, which can move planets out of the bin by reducing their radii below $1.8~R_\oplus$. Atmospheric stripping is strongly time-dependent. Between 50 and 500~Myr, the number of stripped cores increases substantially, with an additional $\sim24\%$ of the full population entering the stripped core pile-up near $X_{\rm atm}\simeq 10^{-4}$. In contrast, between 500 Myr and 3 Gyr, the stripped core population grows only modestly, with only an additional $\sim5\%$ of the population becoming stripped cores. Thus, most of the atmospheric stripping occurs within the first few hundred Myr, when planets are still inflated and the stellar XUV luminosities are high. This behavior is consistent with previous photoevaporation studies \citep[e.g.,][]{2017ApJ...847...29O, 2021MNRAS.503.1526R}, although our calculation further couples it to tidal migration and thermal contraction.

We next quantify this evolution in terms of occurrence rates. The sub-Neptune and super-Earth occurrence rates are computed as the percentage of the full $\sim 20{,}000$ planet sample that lies within each corresponding period-radius bin. 
We note that this estimate is imperfect since we have not accounted for planet multiplicity, the underlying primordial planet distribution, and the fact that not all planets are expected to undergo efficient ZLK-driven HEM (refer to \S~\ref{sec:caveats} for further discussion). Our model does not capture these effects and can only provide an upper-limit estimate of the contribution that the HEM channel makes to the evolution of small-planet occurrence rates. We therefore focus primarily on the age-dependent trends rather than on matching the absolute observed occurrence rates. 
\begin{figure}[!t]
    \centering
    \includegraphics[width=\linewidth]{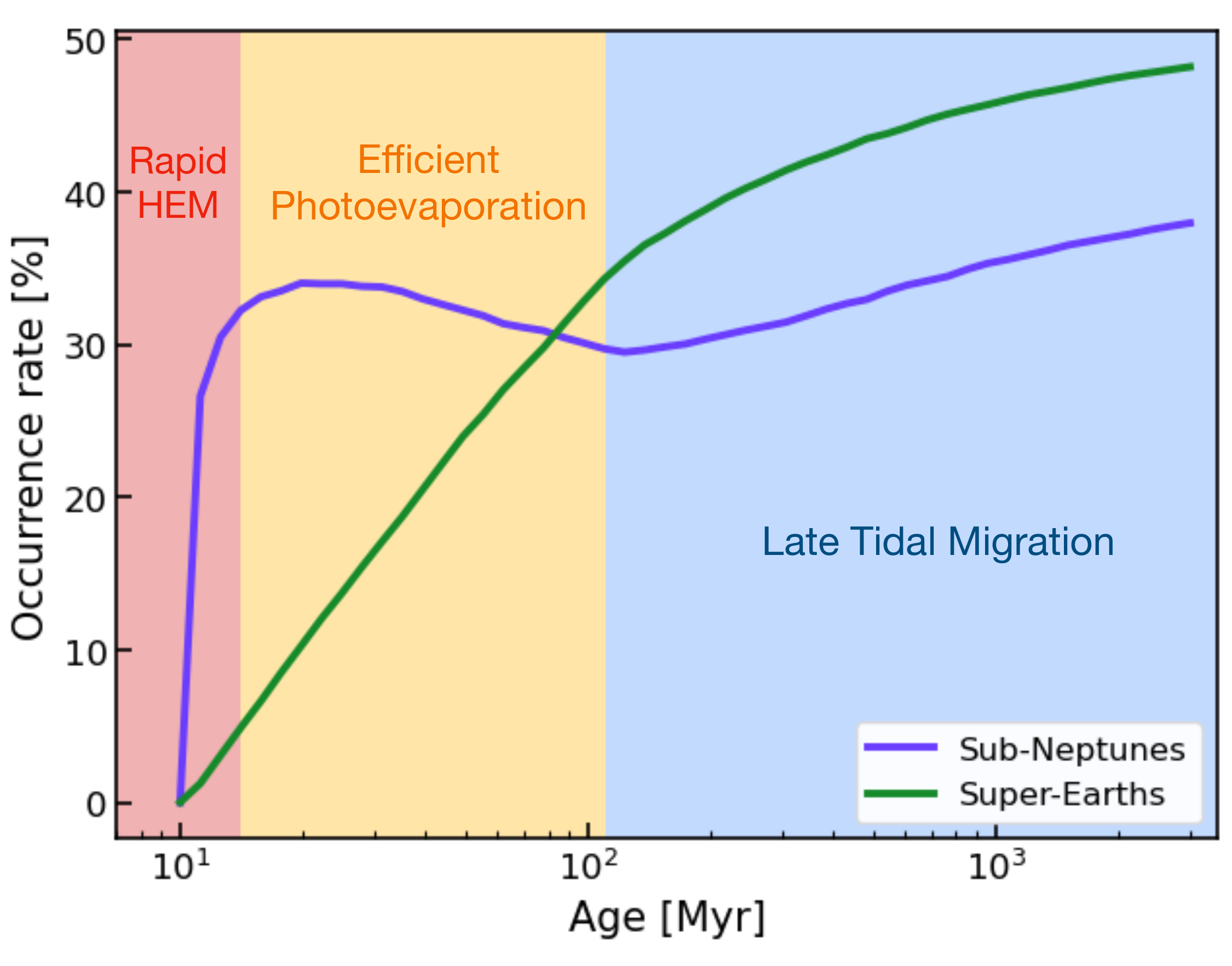}
    \caption{Occurrence-rate evolution of sub-Neptunes and super-Earths from the coupled evolution model. The shaded regions indicate three qualitative phases: rapid HEM, which delivers planets to short periods within $\lesssim 15$~Myr; efficient photoevaporation, during which atmospheric stripping reduces the sub-Neptune occurrence rate and increases the super-Earth occurrence rate; and late tidal migration, where some planets continue to enter the close-in bin and atmospheric escape becomes less efficient.}
    \label{fig:3regimes}
\end{figure}
With these caveats in mind, Figure \ref{fig:3regimes} displays the occurrence rate evolution over time. The behavior can be summarized in terms of three temporal regimes:
\begin{enumerate}
    \item \textbf{Rapid HEM} \textbf{phase}: During this phase, HEM very quickly delivers planets to short orbital periods within $\lesssim 15$~Myr, causing the sub-Neptune occurrence rate to rise sharply. This early delivery is a consequence of our model retaining only systems capable of efficient HEM, most of which reach very high eccentricities. Indeed, $> 90\%$ of the population has $e_0>0.9$ at ZLK quenching (see Figure~\ref{fig:planet_distributions}). 
    
    \item \textbf{Efficient photoevaporation phase}: This phase spans $\sim 15$--$110$~Myr. During this interval, additional planets continue to migrate into the sub-Neptune bin, but photoevaporation is also very efficient. Because photoevaporation dominates over tidal migration during this phase, the sub-Neptune occurrence rate declines mildly and the super-Earth occurrence rate continues to rise.
    \item \textbf{Late tidal migration phase}: This phase begins after $\sim 110$~Myr. At these later ages, atmospheric escape becomes less efficient, and only a small fraction of planets are stripped. The occurrence rate evolution is then dominated by the small population of planets that continue to tidally migrate into the close-in bin, producing the modest increase in the sub-Neptune occurrence rate toward Gyr ages.
\end{enumerate} 

\begin{figure*}
    \centering
    \includegraphics[width=\linewidth]{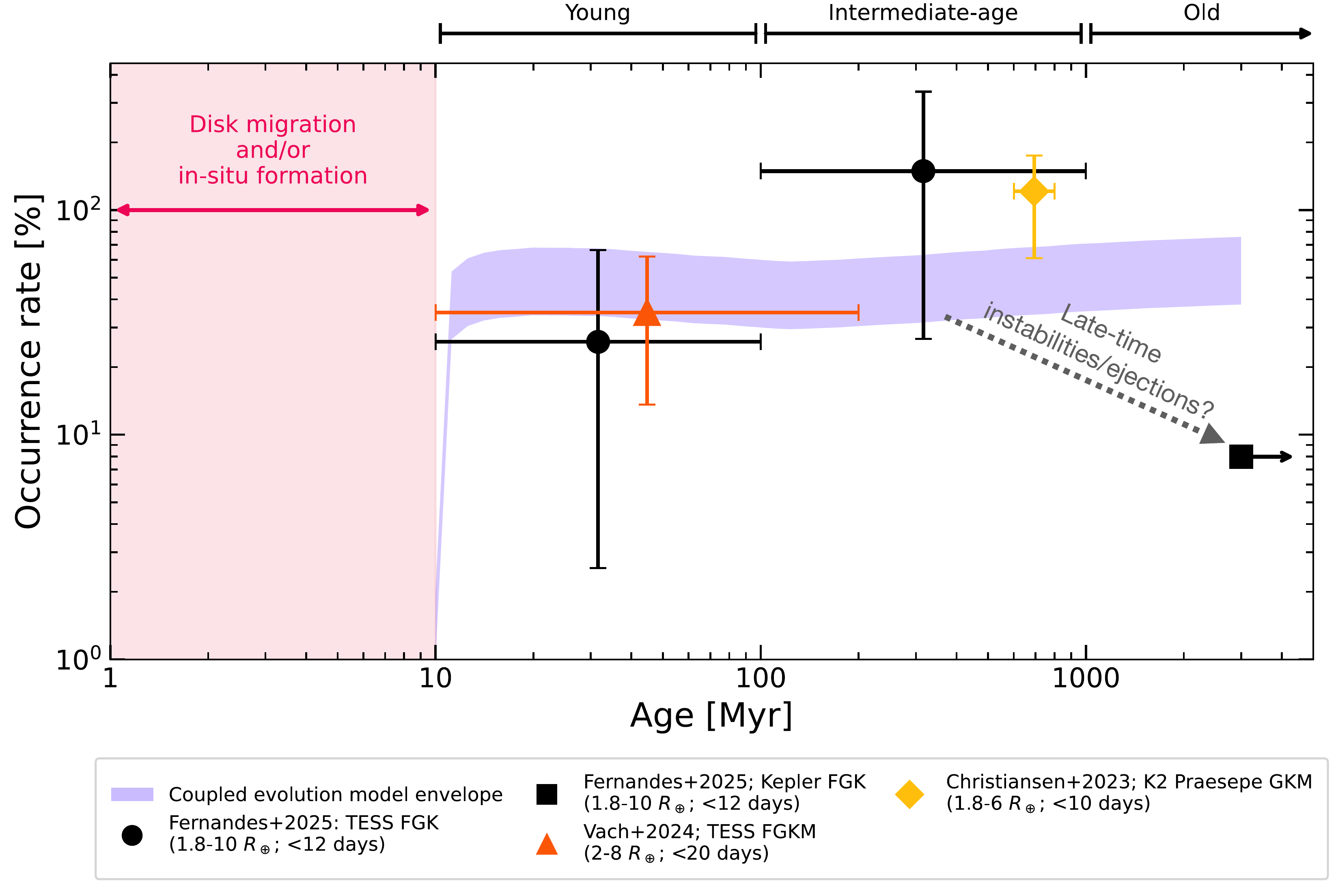}
    \caption{Comparison between the modeled and observed age evolution of the sub-Neptune occurrence rate. The purple shaded region shows the multiplicity-scaled model envelope from our coupled evolution model. Observational measurements from \citet{2025AJ....169..208F}, \citet{2024AJ....167..210V}, and \citet{2023AJ....166..248C} are shown for comparison, with the corresponding period-radius cuts listed in the legend. Error bars denote $2\sigma$ uncertainties.}
    \label{fig:obs_comparison}
\end{figure*}

As a sanity check, we tested the inclusion of core-powered mass loss, a possible late-time atmospheric escape channel \citep{2018MNRAS.476..759G, 2019MNRAS.487...24G}, on a representative subset of $\sim 2000$ systems. The final period-radius distribution is broadly unchanged in this test, suggesting that the population-level trends reported here remain robust even though our coupled evolution model includes photoevaporation as the only atmospheric escape channel. We find that in systems where core-powered mass loss would be most effective, photoevaporation is typically even stronger and therefore dominates the atmospheric escape, while planets that retain H/He envelopes generally have larger core masses and correspondingly slower core-powered mass-loss rates. As a result, adding core-powered mass loss produces only minor additional changes to the occurrence rate evolution. This is consistent with recent studies finding that core-powered mass loss has a limited impact on small-planet demographics at late times \citep{2021MNRAS.503.5658M, 2024MNRAS.529.2716R, 2024ApJ...976..221T, 2024MNRAS.528.1615O}.

\subsection{Comparison with Observations}

The main goal of this paper is to compare our modeled occurrence rate trends with the observations from \cite{2025AJ....169..208F}, who identified two possible age-dependent features. First, they found that the occurrence rate of sub-Neptunes rises from $25.89^{+20.18}_{-11.67}\%$ for young ($10$--$100$~Myr) stars to $148.93^{+93.60}_{-61.10}\%$ for intermediate-age ($100$~Myr--$1$~Gyr) stars, although this increase remains consistent with no change within $2\sigma$ uncertainties. These estimates are broadly consistent with other independent measurements of young \citep{2024AJ....167..210V} and intermediate-age planets \citep{2023AJ....166..248C}. Second, \citet{2025AJ....169..208F} find a much lower occurrence rate, $7.98^{+0.37}_{-0.35}\%$, for the old ($\gtrsim 1$~Gyr) Kepler FGK sample, suggesting a decline from intermediate to Gyr ages. 

While these results were some of the first attempts to measure the evolution of short-period planet occurrence across young and intermediate ages, the current observational constraints remain limited by small-number statistics. In particular, \citet{2025AJ....169..208F}'s occurrence rate estimates were based on only 11 planets within their adopted period, radius, and stellar-type cuts, with four planets in the 10–100 Myr bin and seven in the 100 Myr–1 Gyr bin. The occurrence rates were derived using TESS observations of nearby young clusters and moving groups, and, unlike Kepler, TESS was not designed as a homogeneous occurrence rate survey. 
Moreover, TESS's target selection strategy can introduce additional biases for demographic studies \citep{2021PASP..133i5002F, 2022AJ....163..297C}. Our comparison between the modeled and observed age trends should therefore be regarded as preliminary, motivating larger and more uniformly selected samples of young planets.

With these observational limitations in mind, the measurements from \cite{2025AJ....169..208F} are compared with our coupled evolution model in Figure~\ref{fig:obs_comparison}. We note again that this comparison is intended to test whether the model reproduces the two proposed age-dependent features: the rise of sub-Neptune occurrence rates from young to intermediate ages and the subsequent decline from intermediate to old ages, rather than comparing the exact values. Because the observed occurrence rates are reported as planets per star, while our model evolves only one planet per star, we show a multiplicity-scaled model envelope whose lower edge corresponds to the raw model prediction, and the upper edge shows the same prediction scaled upward by a factor of two to illustrate an approximation of the possible contribution from multiple-planet systems. Further discussion of this correction is provided in Appendix~\ref{sec:occ_rate_normalization}. We organize the comparison between data and model into two sections as follows. 

\subsubsection{Young to Intermediate Age Evolution} \label{subsec:young_to_int}

After the early rapid HEM phase, our model predicts a nearly flat evolution in the sub-Neptune occurrence rate, with variations within $\sim \pm6\%$. Thus, our coupled evolution model does not predict a rise from young to intermediate ages, unlike \cite{2025AJ....169..208F} observed. However, due to the large observational uncertainties, an approximately flat curve could still be consistent with the data from young and intermediate ages within the current $2\sigma$ limits. 

\subsubsection{Intermediate to Old Age Evolution} \label{subsec:int_to_old}
The main discrepancy appears when comparing the intermediate-age population with the old Kepler population. Our model does not reproduce the substantial decline in occurrence rate observed by \cite{2025AJ....169..208F}. Instead, after the efficient photoevaporation phase, planets cool and contract, photoevaporation weakens, and only a small fraction of planets are stripped. At the same time, residual tidal migration continues to supply a small number of new planets into the sub-Neptune bin. As a result, the model increases slightly toward the Gyr ages. Including core-powered mass loss does not resolve this discrepancy either. 

Thus, the physical processes included in our coupled evolution model are not sufficient to explain the observed decline from intermediate-age to old stellar populations. This conclusion is supported by the population-level limits of the HEM fraction discussed in Appendix~\ref{sec:occ_rate_normalization}, suggesting that HEM is not the only pathway shaping the occurrence-rate evolution of close-in sub-Neptunes. Additional formation or dynamical pathways are required, including potentially multi-planet dynamics, late-time dynamical instabilities, and possible differences between cluster and field-star populations. These considerations will be addressed further in Section \ref{sec:caveats}. At the same time, more homogeneous occurrence rate measurements across young, intermediate, and old stellar populations will be important for placing stronger constraints on the true age evolution.

\subsection{Evolution of the Radius Valley}
\begin{figure}[t]
    \centering
    \includegraphics[width=\linewidth]{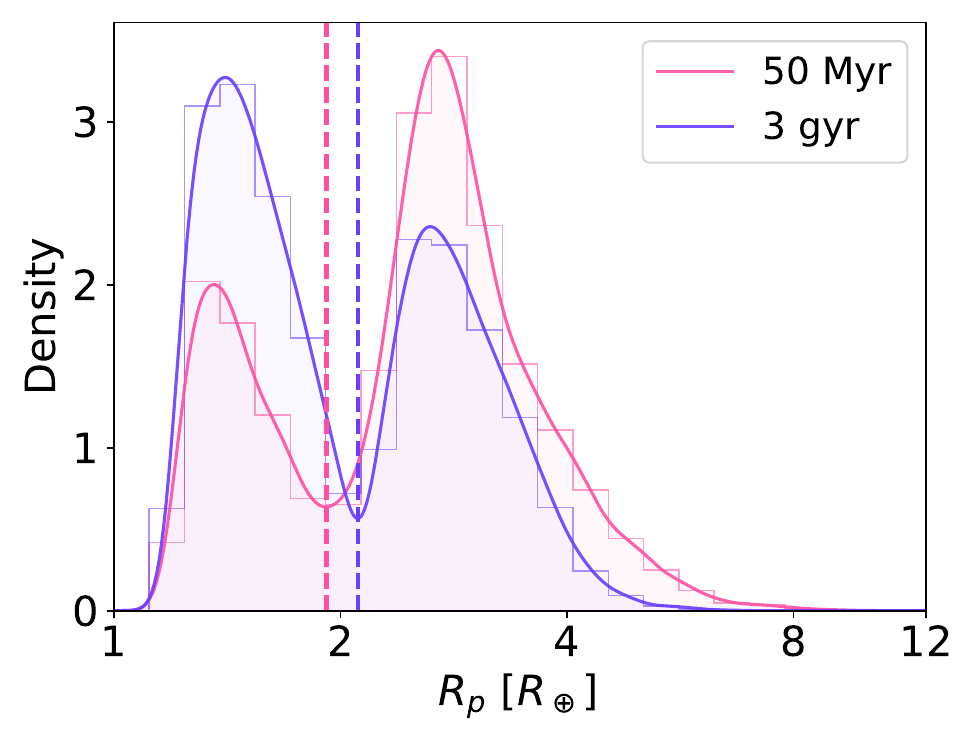}
    \caption{Modeled radius distributions at 50~Myr (pink) and 3~Gyr (purple). The step histograms show the binned distributions, while solid curves show Gaussian KDEs. Dashed vertical lines mark the KDE-inferred radius-valley or the minima between the two peaks.}
    \label{fig:rad_valley_dist}
\end{figure}
Another angle to explore our results is the comparison between the modeled and observed time evolution of the radius valley. \citet{2021AJ....161..265D} analyzed the CKS sample and found that the radius valley shifts to larger radii with increasing stellar age. 
Our coupled evolution model reproduces this behavior qualitatively, as shown in Figure~\ref{fig:rad_valley_dist}, where we compare the radius distributions from our model at 50~Myr and 3~Gyr. The radius valley shifts toward slightly larger radii, while the sub-Neptune peak decreases and the super-Earth peak grows. This behavior indicates that a fraction of initially envelope-bearing sub-Neptunes lose their atmospheres and move into the super-Earth population. Planets stripped at later ages tend to have more massive cores than those stripped early. Their stripped remnants therefore occupy somewhat larger radii within the super-Earth population, shifting the valley between the two peaks toward larger radii. We do not expect an exact quantitative match to the observed CKS radius distribution since our population represents only the subset of systems undergoing efficient ZLK-driven HEM and does not include the full primordial population of super-Earths and sub-Neptunes.

\section{Discussion} \label{sec:discussion}
\subsection{Observational Predictions for Super-Earths} \label{sec:predictions_for_SE}
Our analysis has so far focused on sub-Neptunes, but our coupled evolution model also has predictions for the super-Earth occurrence rate. Observationally, \citet{2022AJ....164..190B} find that the old Kepler super-Earth occurrence rate is higher than the corresponding sub-Neptune occurrence rate. Our model reproduces this qualitative behavior. At $\gtrsim 1$~Gyr ages, the super-Earth occurrence rate exceeds the sub-Neptune occurrence rate. 

The qualitative agreement at old ages motivates us to use the model to make predictions for younger and intermediate-age super-Earth populations. Because the model curve depends on the efficiency of the HEM channel, the results should be interpreted with caution. Sensitivity tests discussed in Appendix~\ref{sec:occ_rate_normalization} show that reducing the HEM efficiency lowers the normalization of the occurrence-rate curves without substantially altering their general temporal evolution. Therefore, the relative occurrence rate evolution of sub-Neptunes and super-Earths provides a useful metric for observational prediction since it is less sensitive to the uncertain normalization of the HEM channel.

There are two other caveats with our super-Earth occurrence rates. First, our calculation does not include a primordial super-Earth population that may already exist after disk dispersal \citep[e.g.,][]{2013MNRAS.431.3444C, 2021A&A...650A.152I}. Second, our synthetic population is restricted to ${M_{\rm core}\in[2,20]~M_\oplus}$, since the primary goal of this study is to model the occurrence-rate evolution of sub-Neptunes. This choice excludes lower-mass cores that would be more easily stripped after migration and could therefore contribute substantially to the super-Earth population. In additional tests, we find that planets with $M_{\rm core}\lesssim 1 M_\oplus$ always become super-Earths after migration, while a large fraction of planets with $1<M_{\rm core}/M_\oplus<2$ are also stripped, although a small number can remain sub-Neptunes.

In the synthetic population, sub-Neptunes initially dominate the super-Earths. At the end of rapid HEM phase, the super-Earth occurrence rate is lower than the sub-Neptune occurrence rate by approximately $26.5\%$. However, since the predicted super-Earth occurrence rate is a lower limit, we cannot confidently predict that the true young super-Earth occurrence rate must initially be lower than the sub-Neptune occurrence rate. The more robust prediction is the subsequent growth of the super-Earth population. Atmospheric stripping increases the super-Earth occurrence rate, especially during the efficient photoevaporation phase, and the super-Earth occurrence rate overtakes the sub-Neptune occurrence rate by $\sim 90$~Myr. Since including primordial super-Earths or lower-mass cores would only increase the super-Earth occurrence rate, this transition should occur by $\sim 90$~Myr at the latest.

Thus, in our model, the combined effects of HEM and photoevaporation
predict that (i) the super-Earth occurrence rate should rise from young to intermediate ages, and (ii) the super-Earth occurrence rate should exceed the sub-Neptune occurrence rate by $\sim 90$~Myr at the latest. HEM-delivered super-Earths may also retain distinctive architectures: tides may erase their eccentricities but stellar spin--orbit misalignments may persist for longer ages. Residual eccentricities or stellar obliquities could therefore provide tentative evidence that individual super-Earths were delivered through this channel. 

Observationally testing our occurrence rate predictions will require better constraints on young super-Earths. Such measurements remain difficult because young stars are rapidly rotating \citep{2020MNRAS.496.1197B, 2026MNRAS.547ag454S}, active \citep{2020AJ....160..219F}, spotted, and frequently variable \citep{2022MNRAS.511.4285B, 2024MNRAS.529.4442S}, making shallow transits from small planets harder to detect and validate. Existing young planet occurrence studies have therefore focused mainly on larger planets \citep[e.g.,][]{2024AJ....167..210V, 2025AJ....169..208F, 2023AJ....166..248C}. However, recent discoveries of small planets around young stars show that this regime is becoming increasingly accessible \citep[e.g.,][]{2024AJ....167...54C, 2025AJ....170...32B}. Future measurements of young super-Earth occurrence rates will therefore provide a direct test of our prediction.

\subsection{Tidally Induced Radius Inflation} \label{sec:tidal radius inflation}
\begin{figure}
    \centering
    \includegraphics[width=\linewidth]{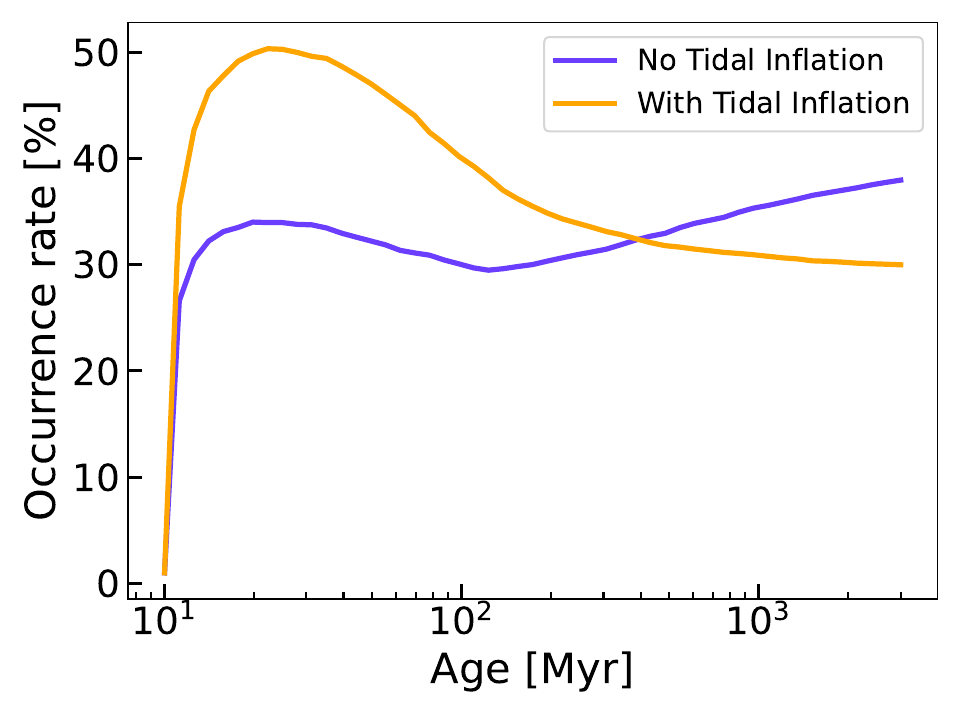}
    \caption{Effect of tidally induced radius inflation on the sub-Neptune occurrence-rate evolution. The purple and orange curves show the coupled evolution model without and with tidal inflation, respectively.}
    \label{fig:tidal_inflation}
\end{figure}

Tidally induced radius inflation is an important additional effect to consider when modeling HEM-driven occurrence-rate evolution. For eccentric planets on short-period orbits, time-varying distortions in the planet's shape dissipate orbital energy as heat inside the planet \citep{1981A&A....99..126H, 2008ApJ...681.1631J, 2010A&A...516A..64L}. This heating can significantly inflate planetary radii, in some cases by factors of order $\sim 2$ for Neptune-size planets \citep[e.g.,][]{2001ApJ...548..466B, 2007A&A...462L...5L, 2019ApJ...886...72M, 2020ApJ...897....7M, 2025ApJ...979..218L, 2025ApJ...988..247S, 2026ApJ..1003...84I}.

This effect is especially relevant because both tidal migration and atmospheric mass loss depend sensitively on $R_p$. In our coupled evolution model, $R_p$ evolves through thermal cooling and atmospheric mass loss, but we haven't included a tidal heating contribution to the planet's internal energy budget. To assess the possible impact of tidally induced radius inflation, we run a modified version of the coupled evolution model that accounts for tidal heating. The details of this implementation are described in Appendix~\ref{sec:tidal_heating_modeling}.

Because most planets in our synthetic population begin on highly eccentric orbits with small pericenter distances, unrestricted tidal dissipation can drive runaway inflation \citep{2026ApJ...997..138H, 2026ApJ...997..139H}. If the dissipated energy is efficiently deposited in the deep convective zone, many planets rapidly lose their H/He envelopes and become stripped cores. Since modeling tidal heat deposition and transport is beyond the scope of this work, we adopt a phenomenological cap on the tidal luminosity deposited at the deep convective zone, assuming all the excess tidal power is radiated away. The sensitivity of the occurrence rate evolution to the adopted tidal luminosity cap is explored in Appendix~\ref{sec:runaway inflation}. 
We therefore treat this as an exploratory calculation, aimed at assessing how moderate tidally induced radius inflation could qualitatively affect the occurrence-rate evolution.


We compare the occurrence rate evolution with and without tidally induced radius inflation in Figure~\ref{fig:tidal_inflation}. Including tidal inflation changes both the normalization and the shape of the sub-Neptune occurrence-rate evolution. Tidal heating inflates the planetary radius, and larger radii enhance tidal dissipation, allowing more systems to migrate rapidly into the close-in bin during the early rapid HEM phase. As a result, the early occurrence rate peak derived from the tidal inflation model is $\sim 12\%$ higher compared to that derived from the no inflation model. The rapid HEM phase also lasts slightly longer when tidal inflation is included, extending to $\sim 25$~Myr compared to $\sim 15$~Myr in the model without tides.

The subsequent evolution is also modified. Inflated planets have larger radii and more weakly bound envelopes, making photoevaporative mass loss more efficient. As a result, the decline after the rapid HEM phase is stronger and more extended when tidal inflation is included. Without tidal inflation, the drop in the sub-Neptune occurrence rate ends by $\sim 110$~Myr, with a total drop of only $\sim 5\%$. With tidal inflation, planets remain susceptible to atmospheric stripping for longer, so the decline continues until $\sim 500$~Myr and decline by $\sim 19\%$. 

Both models remain consistent with the large observational uncertainties reported by \citet{2025AJ....169..208F}. However, neither model supports the interpretation that the sub-Neptune occurrence rate might rise gradually from young to intermediate ages because of continued tidal migration. In both cases, the migration-driven sharp rise occurs early on at very young ages, following which the no inflation model is roughly flat whereas the tidal-inflation model declines because mass loss becomes more efficient.

Moreover, including tidal inflation still does not reproduce the sharp decline in occurrence rates from intermediate to old ages observed by \citet{2025AJ....169..208F}. In the no inflation case, the occurrence rate increases slowly toward Gyr ages due to continued tidal migration of a small residual population. With tidal inflation, this late-time increase in occurrence rate is suppressed and it remains approximately flat. This occurs because radius inflation enhances tidal dissipation which reduces the circularization timescales causing most planets in our sample to reach the close-in bin during the rapid HEM phase. At the same time, most of the atmospheric stripping occurs within the first $\sim 500$~Myr. At later ages, planets have largely circularized, tidal heating has weakened, radii have contracted, and photoevaporation has become less efficient. As a result, the model neither strongly replenishes the sub-Neptune population nor removes enough planets from it to produce a substantial decline. Additionally, because tidal inflation enhances early migration and also strengthens subsequent atmospheric stripping, the occurrence rate at 3~Gyr from the tidal inflation model is slightly lower than in the no inflation case. 

\subsection{Model Caveats and Future Extensions} \label{sec:caveats}
Several ingredients not included in our coupled evolution model could affect the occurrence rates and their time evolution. First, our model assumes one planet per star and allows every modeled planet to undergo HEM. While growing evidence indicates a prevalence of HEM in close-in Neptunes and sub-Neptunes \citep[e.g.][]{2023A&A...669A..63B, 2024ApJ...972..159Y, 2024A&A...689A.250C, 2025ApJ...979..218L, 2026A&A...709L..17C}, simultaneously, a substantial fraction of close-in planets ranging from super-Earth to Neptune sizes reside in compact multis \citep{2011ApJS..197....8L, 2014ApJ...784...45R, 2018ApJ...860..101Z}. These systems often exhibit correlated sizes and orbital spacings characteristic of ``peas-in-a-pod" architectures \citep{2017ApJ...849L..33M, 2018AJ....155...48W}. Such compact architectures cannot be reconciled with large-amplitude ZLK oscillations followed by HEM \citep{2008ApJ...683.1063T, 2021ApJ...923..118W, 2022ApJ...932...78F}, and other formation pathways must explain them \citep[e.g.][]{2017MNRAS.470.1750I}. This is particularly relevant for the occurrence-rate measurements considered here since more than half of the planets in the \citet{2024AJ....167..210V, 2025AJ....169..208F} samples belong to known multi-planet systems. Our model should therefore be interpreted as describing only the subset of systems in which HEM remains dynamically viable. 

Compact multis can also undergo dynamical evolution that is absent from the present calculation. Resonant coupling \citep{2014ApJ...790..146F}, secular interactions \citep{2011ApJ...735..109W}, planet--planet scattering \citep{2008ApJ...686..580C}, and ejections \citep{1996Sci...274..954R} can alter both the surviving planet multiplicity and the orbital distribution, modifying the normalization and potentially the time dependence of the occurrence rate curve independently of HEM. The multiplicity correction used to illustrate the model envelope in Figure~\ref{fig:obs_comparison} is also only an approximate normalization, not a self-consistent model treatment of compact multi-planet dynamics. 

There is no obvious reason, however, for compact multi dynamics to produce the observed rise in sub-Neptune occurrence from young to intermediate ages since these systems are thought to be emplaced during the disk lifetime (either by in situ formation or disk migration) \citep[e.g.][]{2017MNRAS.470.1750I}.
On the other hand, delayed dynamical instabilities at later times could reduce the number of surviving close-in planets through collisions, mergers, or ejections and potentially contribute to the decline from intermediate to old ages \citep{2020PNAS..11718194T, 2026A&A...707A.285G}. 
We will explore their potential contribution to the late-time occurrence-rate decline in future work. 


Another limitation of our model is the simplified planet interior structure comprising Earth-like cores and H/He envelopes. Real sub-Neptunes must span a broader range of compositions, including volatile-rich interiors and more complex envelope structures \citep[e.g.,][]{2021JGRE..12606639B, 2022Sci...377.1211L, 2024A&A...688A..59P}. These differences can affect the mass-radius relation, atmospheric binding energy, cooling history, and therefore the occurrence-rate evolution. In addition, we model photoevaporation using a simple energy-limited approach, which may not fully capture the hydrodynamics of strongly irradiated close-in sub-Neptunes \citep[e.g.,][]{2026ApJ..1008...11S}. Future work could explore more realistic hydrodynamic models, such as Wind-AE, that remain computationally tractable for population-level studies \citep{2025ApJ...995..198B}.

Finally, our model does not specify the primordial close-in planet population after disk dispersal. Some super-Earths and sub-Neptunes may already be present at this stage through in-situ formation, or disk-driven migration \citep[e.g.,][]{2013MNRAS.431.3444C, 2021A&A...650A.152I}. Therefore, the results presented in this paper only tell us about the contribution of the HEM channel to sub-Neptune occurrence rates, rather than the total close-in small-planet occurrence rate.

\section{Conclusions} \label{sec:conclusions}
Motivated by observations from \cite{2025AJ....169..208F} of sub-Neptune occurrence rates over time, we set out to test the hypothesis that tidal migration and mass loss drive their evolution. We developed a unified framework that couples thermal cooling, atmospheric mass loss, and orbital evolution in close-in small planets. We created a synthetic population of systems and subjected them to efficient ZLK-driven high-eccentricity migration, evolving them from their post-ZLK quenching time to 3~Gyr. We used this framework to estimate the maximum contribution that HEM can make to the age evolution of close-in small planets and compared the results with observed age-dependent trends. Our main results are summarized below.
\begin{enumerate}
\item The coupled evolution model qualitatively reproduces key demographic features of the mature planet population, including the radius valley and the hot-Neptune desert. 

\item The occurrence-rate evolution proceeds through three temporal regimes. In the first $\sim15$~Myr, a rapid HEM phase delivers many planets to short orbital periods, producing a sharp rise in the sub-Neptune occurrence rate. The efficient photoevaporation phase from $\sim 15$ to $110$~Myr reduces the occurrence rate of sub-Neptunes while increasing that of super-Earths. Finally, a late tidal migration phase delivers a residual population of planets inwards.

\item 
Between 100~Myr and 1~Gyr, the occurrence rate increases by only $\sim 5\%$. Given the current $2\sigma$ observational uncertainties \citep{2025AJ....169..208F}, such weak evolution remains compatible with the trends that could be inferred from the data, but it does not favor a scenario in which the occurrence rate rises from young to intermediate ages due to sustained tidal migration.

\item The model does not reproduce the sharp decline in sub-Neptune occurrence rate from intermediate to old ages observed by \citet{2025AJ....169..208F}. Instead, during this time, photoevaporation becomes less efficient, while residual tidal migration continues to deliver a small number of planets into the close-in bin. Including core-powered mass loss does not resolve this discrepancy. The modeled occurrence rate increases mildly toward Gyr ages. In future work, additional physics will be needed to explain the decline in sub-Neptune occurrence rate from intermediate to old ages. Possibilities include multi-planet dynamics, late-time instabilities, and possible differences between cluster and field-star populations. 

\item The model predicts that the radius valley evolves with age. As atmospheric stripping proceeds, the valley shifts towards larger radii, qualitatively consistent with the observational trends reported by \citet{2021AJ....161..265D}.

\item If the HEM channel contributes significantly to the close-in small-planet population, we predict a rising super-Earth occurrence rate from young to intermediate ages, with the super-Earth occurrence rate exceeding the sub-Neptune occurrence rate by $\sim 90$~Myr at the latest.

\item 
Including radius inflation from tidal heating produces a higher early occurrence rate peak than in the no-inflation model and extends the rapid HEM phase to $\sim 25$~Myr. Additionally, because the atmospheric mass loss is more efficient, the post-peak decline becomes stronger and lasts until $\sim 500$~Myr. However, even with tidal inflation, the occurrence remains roughly flat at Gyr ages and still does not reproduce the observed decline from intermediate to old ages.
\end{enumerate}

Looking ahead, the most direct observational test of this picture will come from improved occurrence-rate measurements of young and intermediate-age small planets, especially super-Earths. Efforts such as the THYME collaboration are already expanding the sample of young transiting planets \citep{2019ApJ...880L..17N, 2021AJ....161...65N, 2021AJ....161..171T, 2024AJ....167...54C}, but detecting small planets around young stars remains challenging. Improved light-curve modeling, stellar-activity mitigation, and targeted searches in young clusters will therefore be crucial for testing our predictions. Complementary constraints on planets during or shortly after the disk phase would also help determine how much of the small-planet population is primordial, rather than produced later by migration.

\section*{Acknowledgements}
R.S. gratefully acknowledges Ilaria Pascucci, Galen Bergsten, Tiger Lu, and Utkarsh for many helpful conversations. This material is based upon work supported by the National Science Foundation under grant No. 2306391. R.S. acknowledges the support from the Quad Fellowship by International Institute of Education. S.M. acknowledges support from the Alfred P. Sloan Foundation through a Sloan Research Fellowship.
\appendix
\section{Occurrence-Rate Normalization} \label{sec:occ_rate_normalization}
Observed occurrence rates are commonly reported as the number of planets per star in a specified period-radius bin. These rates can therefore exceed 100\% when multiple planets per star contribute to the same bin \citep[e.g.,][]{2022AJ....164..190B, 2023AJ....166..248C, 2024AJ....167..210V, 2025AJ....169..208F}. Because our coupled evolution model only follows one planet per star, the modeled occurrence rate cannot exceed 100\% by construction. The raw model curve in Figure~\ref{fig:3regimes} should thus be interpreted as the fraction of our simulated population occupying the sub-Neptune bin, rather than as a fully multiplicity-corrected number of planets per star.

To make an approximate comparison with observed occurrence rates, the modeled curves in Figure~\ref{fig:obs_comparison} contain an envelope that represents the multiplicity correction. The lower edge of the envelope corresponds to the raw one planet per star prediction from our coupled evolution model, while the upper edge rescales this prediction upward by a factor of two. This rescaling was not derived with a detailed population synthesis, but nevertheless it represents an optimistic upper limit motivated by the fact that multi-planet systems contribute substantially to the observed small-planet population (e.g. \citet{2018ApJ...860..101Z} infer an average of $3.0\pm0.3$ planets per star within 400 days). Kepler demographics also show that roughly 40\% of detected transiting planet candidates reside in multi-planet systems \citep{2011ApJS..197....8L, 2014ApJ...784...45R}. 

\begin{figure}
    \centering
    \includegraphics[width=\linewidth]{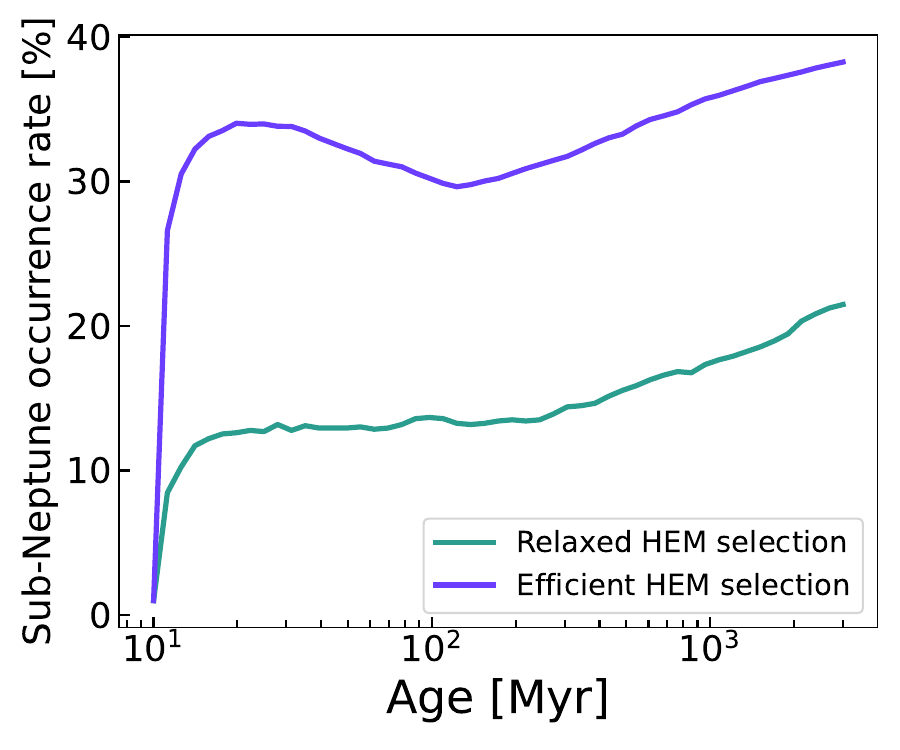}
    \caption{Sensitivity of the modeled sub-Neptune occurrence rate to the HEM selection criteria. Purple and green curves show the age evolution of the sub-Neptune occurrence rate for the efficient-HEM and relaxed-HEM populations, respectively. The efficient-HEM population uses $t_{\rm ZLK}<100$~Myr and $\psi_{bc}$ concentrated near $90^\circ$, while the relaxed-HEM population removes the timescale cut and draws $\psi_{bc}$ uniformly across the ZLK-active window.}
    \label{fig:sensitivity_test}
\end{figure}

Alongside the multiplicity considerations, the modeled occurrence rates are also affected by the efficiency of the HEM channel. The synthetic planet population was constructed to isolate systems capable of efficient ZLK-driven HEM, using cuts such as $t_{\rm ZLK} < 100$~Myr and $\psi_{bc}$ concentrated near $90^\circ$ (see \S~\ref{sec:pop_synthesis}). To test the sensitivity of our results to these cuts, we repeated the evolution for a representative subset of $\sim1500$ systems after removing the $t_{\rm ZLK}$ timescale cut and drawing $\psi_{bc}$ from a uniform distribution across the ZLK-active window. The resulting occurrence-rate evolution is shown in Figure~\ref{fig:sensitivity_test}

Relaxing these selections substantially lowers the overall sub-Neptune occurrence rate because a smaller fraction of systems reach
high eccentricities needed for efficient HEM. 
The slight decline during the efficient photoevaporation phase described in \S~\ref{subsec:timeevol} is also less pronounced, because fewer planets are delivered to close-in orbits where photoevaporation can efficiently strip their envelopes. However, the general time evolution remains similar, featuring an initial rapid rise, a modest decrease or flattening during the efficient photoevaporation phase, and a subsequent increase toward $\sim$Gyr ages as tidal migration continues to deliver planets. In particular, removing the selection cuts does not produce either a sustained rise from young to intermediate ages or a decline from intermediate to old ages as suggested by observations. Thus, while the HEM-efficiency selections primarily affect the normalization of the sub-Neptune occurrence rate curve and the strength of the early photoevaporation feature, the main age-dependent trends discussed in this work remain unchanged.


More generally, the contribution of the HEM channel to the occurrence rate evolution of sub-Neptunes can be written schematically as, 
\begin{equation}
    f_{\rm subNep} \simeq f_{\rm HEM}f_{\rm sim}(t)
\end{equation}
where $f_{\rm sim}(t)$ is the simulated occurrence rate from our coupled evolution model. The factor, $f_{\rm HEM} \in (0, 1]$ represents the fraction of systems for which HEM is effective. This expression is only an approximate normalization, since changing the HEM efficiency does not perfectly rescale the occurrence-rate curve, although its broad time-dependent behavior remains similar, as shown in the sensitivity tests above. Because our synthetic population is constructed using physical and timescale cuts that preferentially select systems capable of efficient HEM, it corresponds to the $f_{\rm HEM} \simeq 1$ limit. Therefore, the model envelope in Figure~\ref{fig:obs_comparison} should be interpreted as the maximum contribution that the ZLK-driven HEM channel could make to the sub-Neptune occurrence rate, within the assumptions of our coupled evolution model. To match the model with the observed occurrence rates within the uncertainty limits, a relatively large HEM efficiency factor, $f_{\rm HEM} \gtrsim 0.45$ is required. This estimate is only illustrative, since the model does not include a primordial short-period planet population formed or delivered during the protoplanetary disk phase. 

It is challenging to estimate the true value of $f_{\rm HEM}$. As for the mechanism explored in this work -- ZLK-driven migration with a stellar binary -- a crude population-level estimate suggests a small HEM contribution. Roughly 50\% of stars reside in multiple stellar systems \citep{2013ARA&A..51..269D}, and in our population synthesis only $\sim$10\% of the initially drawn systems satisfy the ZLK-timescale, and Roche-survival requirements adopted for efficient HEM (see \S~\ref{subsec:cuts}). Combining these factors gives $f_{\rm HEM} \sim 0.05$ \citep[e.g.][]{2016MNRAS.456.3671A}. However, there are other drivers of HEM, including planet--planet ZLK oscillations, planet--planet scattering, and secular chaos. Quantifying the combined efficiency of all such channels is nontrivial.

Nevertheless, observational evidence suggests that HEM is a common formation channel. It is firmly established to be a major source of the hot Jupiter population \citep{2023PNAS..12004179C, 2026AJ....171..157S}, and the growing evidence is pointing to it being an important source of hot Neptunes as well \citep[e.g.][]{2023A&A...669A..63B, 2025A&A...701A.190B, 2026A&A...709L..17C}. Thus, while stellar-driven ZLK can only account for $f_{\rm HEM} \sim 0.05$ and the total value including all channels is certainly less than 1, it still likely sums to a non-negligible fraction.
\section{Including Tidal Heating in the Coupled Evolution Model} \label{sec:tidal_heating_modeling}
To explore the role of tidally induced radius inflation, we add tidal heating as an additional luminosity source in the entropy evolution equation (Eq.~\ref{eq:dSdt}),
\begin{equation} \label{eq:dSdt_tides}
    \frac{dS}{dt} = \frac{-L + L_{\rm tide}}{\int_{\rm conv}{Tdm}}.
\end{equation}
The tidal luminosity, $L_{\rm tide}$, is defined as the rate at which orbital energy is dissipated as heat inside the planet \citep{2010A&A...516A..64L} 
\begin{equation}
    L_{\rm tide} = 2K\left[N_a(e) - \frac{N^2(e)}{\Omega(e)} \right].
\end{equation}
This equation assumes zero planetary obliquity. Using Eq.~\ref{eq:dSdt_tides} in place of the Eq.~\ref{eq:dSdt} in the coupled evolution model couples tidal heating directly to the planet’s radius evolution. The resulting change in $R_p$ then feeds back on both atmospheric mass loss and tidal migration.

\subsection{Runaway Inflation and Luminosity Cap} \label{sec:runaway inflation}
\begin{figure}
    \centering
    \includegraphics[width=\linewidth]{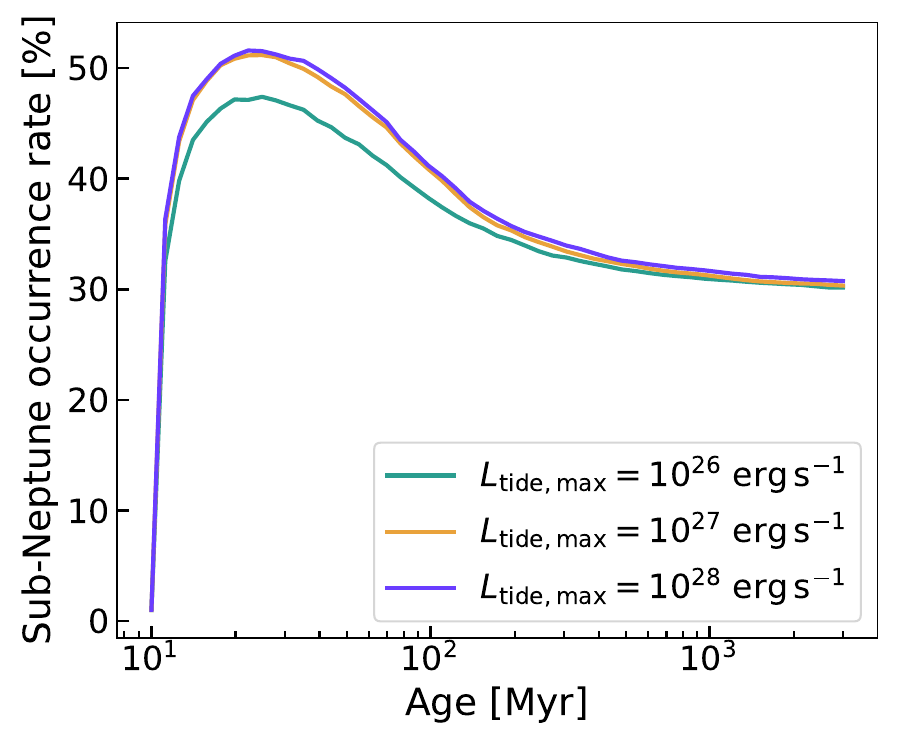}
    \caption{Sensitivity of the sub-Neptune occurrence-rate evolution to the adopted tidal-luminosity cap. Green, orange, and purple curves show results for $L_{\rm tide, \max} = 10^{26}, 10^{27},$ and $10^{28} \rm{ergs} \ \rm{s}^{-1}$, respectively. } 
    \label{fig:lum_cap_sensitivity}
\end{figure}

If the tidal luminosity becomes comparable to or larger than the planet's intrinsic cooling luminosity, $L_{\rm tide}\gtrsim L$, the planet no longer simply cools and contracts. Instead, tidal heating increases the internal entropy and inflates the radius. This mechanism has been invoked to explain inflated, low-density sub-Neptunes and sub-Saturns \citep[e.g.,][]{2019ApJ...886...72M, 2020ApJ...897....7M, 2024Natur.630..836W}. 

Moreover, when tidal heating is coupled to HEM, it can produce a positive feedback \citep{2026ApJ...997..138H}.
This occurs if the dissipated energy is efficiently deposited in the planet's deep convective layers, leading to radius inflation that further increases the tidal dissipation rate. This feedback is especially strong for sub-Neptune-mass planets. In some cases, the planet experiences runaway inflation and cannot retain a stable envelope, and the atmosphere is rapidly lost, leaving behind a stripped core. This runaway inflation problem was seen by \citet{2026ApJ...997..138H, 2026ApJ...997..139H}. If this outcome was typical for HEM-delivered planets, sub-Neptunes would be efficiently converted into super-Earths, making sub-Neptunes rare even at young ages after the rapid HEM phase. This would be difficult to reconcile with the prevalence of young sub-Neptunes.

Runaway inflation introduces a key uncertainty in tidal inflation modeling; the outcome depends not only on $L_{\rm tide}$, but also on where the heat is deposited. If a large fraction of the heat is deposited in shallow radiative layers, it may be radiated away instead of being stored as internal entropy in the deep convective interior, reducing the degree of radius inflation \citep{2026ApJ...997..138H}. While a more advanced treatment of tides is beyond the scope of this work, we avoid runaway inflation by adopting an ad hoc cap on the tidal luminosity, $L_{\rm tide, \max} = 10^{28} \rm{ergs} \ \rm{s}^{-1}$ retained by the deep interior, $L_{\rm tide, eff} = \min(L_{\rm tide}, L_{\rm tide, \max})\ \rm{ergs} \ \rm{s}^{-1}$, assuming that any tidal power above this value is deposited in shallower layers and radiated away. This cap is not intended to represent a definitive physical threshold, but allows us to explore the qualitative impact of moderate tidally induced radius inflation on the occurrence-rate evolution without letting extreme runaway cases  dominate the population. 

We tested the sensitivity of the results to this choice by repeating the calculation with $L_{\rm tide, \max} \in \{10^{26}, 10^{27}\} \ \rm{erg} \ \rm{s}^{-1}$. The occurrence-rate evolution is  basically converged once the cap, $L_{\rm tide, \max} \gtrsim 10^{27}$. Lowering the cap to $10^{26}  \ \rm{erg} \ \rm{s}^{-1}$ weakens the early HEM driven rise, because reduced radius inflation leads to less efficient tidal dissipation and circularization. At later ages, however, the curves converge and the $\sim$Gyr age occurrence rates remain nearly unchanged. Thus, the adopted tidal-luminosity cap mainly affects the strength of the early HEM-driven rise rather than the long-term occurrence rate evolution. This sensitivity test is shown in Figure~\ref{fig:lum_cap_sensitivity}.

\bibliographystyle{aasjournalv7}
\bibliography{main}

\end{document}